\documentclass[a4paper,fleqn,usenatbib]{mnras}

\usepackage{mathptmx}
\usepackage[T1]{fontenc}
\usepackage{pdflscape}
\usepackage{array}
\usepackage{natbib}

\usepackage{graphicx}	% Including figure files
\usepackage{amsmath}	% Advanced maths commands
\usepackage{amssymb}	% Extra maths symbols
\newcommand{\lrsp}{   {Living Reviews in Solar Physics}}

\title[Mass Loss Rate from the Sun]{Mass Loss via Solar Wind and Coronal Mass Ejections During Solar Cycle 23 and 24}

\author[Mishra et al.]{Wageesh Mishra$^{1, 2}$\thanks{Email: m.wageesh30@gmail.com},
Nandita Srivastava$^{3}$, %\thanks{Email: nandita@prl.res.in}
Yuming Wang$^{1, 2}$, %\thanks{Email: ymwang@ustc.edu.cn}
\newauthor
Zavkiddin Mirtoshev$^{4}$, %\thanks{Email: zmirtoshev@gmail.com} \\
Jie Zhang$^{5}$, %\thanks{Email: jzhang7@gmu.edu}
and Rui Liu$^{1}$
\\
$^{1}$CAS Key Laboratory of Geospace Environment, School of Earth and Space Sciences,\\
University of Science and Technology of China, Hefei 230026, People's Republic of China; wageesh@ustc.edu.cn, ymwang@ustc.edu.cn, rliu@ustc.edu.cn\\
$^{2}$ CAS Center for Excellence in Comparative Planetology, Hefei 230026, People's Republic of China  \\
$^{3}$Udaipur Solar Observatory, Physical Research Laboratory, Badi Road, Udaipur 313001, India; nandita@prl.res.in\\
$^{4}$Department of Physics, Samarkand State University, Samarkand-140104, Uzbekistan; zmirtoshev@gmail.com\\
$^{5}$Department of Physics and Astronomy, George Mason University, Fairfax-22030, Virginia, USA; jzhang7@gmu.edu \\
}

\date{Accepted 8 April 2019}

\pubyear{2018}

\begin{document}
\label{firstpage}
\pagerange{\pageref{firstpage}--\pageref{lastpage}}
\maketitle

\begin{abstract}

Similar to the Sun, other stars shed mass and magnetic flux via ubiquitous quasi-steady wind and episodic stellar coronal mass ejections (CMEs). We investigate the mass loss rate via solar wind and CMEs as a function of solar magnetic variability represented in terms of sunspot number and solar X-ray background luminosity. We estimate the contribution of CMEs to the total solar wind mass flux in the ecliptic and beyond, and its variation over different phases of the solar activity cycles. The study exploits the number of sunspots observed, coronagraphic observations of CMEs near the Sun by \textit{SOHO/LASCO}, in situ observations of the solar wind at 1 AU by \textit{WIND}, and \textit{GOES} X-ray flux during solar cycle 23 and 24. We note that the X-ray background luminosity, occurrence rate of CMEs and ICMEs, solar wind mass flux, and associated mass loss rates from the Sun do not decrease as strongly as the sunspot number from the maximum of solar cycle 23 to the next maximum. Our study confirms a true physical increase in CME activity relative to the sunspot number in cycle 24. We show that the CME occurrence rate and associated mass loss rate can be better predicted by X-ray background luminosity than the sunspot number. The solar wind mass loss rate which is an order of magnitude more than the CME mass loss rate shows no obvious dependency on cyclic variation in sunspot number and solar X-ray background luminosity. These results have implications to the study of solar-type stars.  
 
\end{abstract}


\section{Introduction}
\label{intro}
It is now well known that our Sun loses mass through quasi-steady solar wind and coronal mass ejections (CMEs). The idea of solar wind as the radially uniform supersonic outflow of the stream of charged particles from the hot solar corona was first introduced by \citet{Parker1958}. The existence of solar wind was confirmed on its first direct observation from spacecraft \textit{Luna 1} \citep{Gringauz1960}. Since then numerous studies using the solar wind observations in the ecliptic and beyond from a series of spacecraft (\textit{Luna, Venus probe, Explorer, Ranger, Mariner, Pioneers, Helios, Voyagers, ISEE, Ulysses, WIND, SOHO, ACE, Genesis, and STEREO etc.}) have been made \citep{Lopez1985,McComas2000,Richardson2000,Neugebauer2002,Zurbuchen2006,Stverak2009}. Similar to the Sun, other stars have also shown the spectroscopic evidence of a quasi-steady wind (i.e., stellar wind) \citep{DeJager1988,Dudley1992,Kudritzki2000,Puls2008} from their corona. Although several attempts have been made for understanding stellar winds, their properties are not yet well understood, both theoretically and observationally \citep{Parker1960,Lamers1999,Matt2008}. Even for the solar wind, the exact physical mechanism for its heating and acceleration is poorly understood despite the studies for decades using remote observations, in-situ observations, as well as theoretical modeling \citep{Marsch2006,Cranmer2007,Ofman2010}.

CMEs are known to be large-scale expulsion of magnetized plasma structures from closed magnetic field regions on the Sun. They were first detected in the coronagraph images taken in 1971 by NASA's OSO-7 spacecraft \citep{Tousey1973}. However, some definite inferences for the solar wind \citep{Eddington1910,Birkeland1916,Biermann1951} as well as CMEs from the Sun \citep{Chapman1931,Eddy1974} were made decades before their formal discovery. Following OSO-7, a series of spacecraft (\textit{Skylab, Helios, P78-1 Solwind, SOHO, Coriolis, and STEREO etc.}) have observed thousands of CMEs leading to a vast literature \citep{Munro1979,Howard1985,Gosling1993,Hundhausen1999,Gopalswamy2000,Schwenn2006,Vourlidas2010,Chen2011,
Wang2011,Webb2012,Mishra2013,Mishra2017,Harrison2018}. CMEs have been observed to occur often having spatial and temporal relation with solar flares, eruptive prominences \citep{Munro1979,Webb1987,Zhang2001,Gopalswamy2003} and with helmet streamer disruptions \citep{Dryer1996}. Unlike CMEs from the Sun, to observe stellar CMEs are challenging because the close stellar environment cannot be spatially resolved. Although stellar CMEs have not yet been directly detected in Thomson scattered optical light from other stars, it is believed that the extreme X-ray flares observed on stars may be in conjunction with extreme stellar CMEs \citep{Houdebine1990,Wheatley1998,Leitzinger2011,Aarnio2012,Osten2015,Vida2016}. Indeed, the stellar X-ray flare, helmet streamers and prominences observed on T Tauri Stars have shown similarities with those observed on the Sun \citep{Haisch1995,Massi2008}. The CMEs and flares themselves may not be causally related, they both seem to be involved with the reconfiguration of complex magnetic field lines within the corona caused by the same underlying physical processes, e.g., magnetic reconnection \citep{Priest2002,Compagnino2017}. But, even for the sun, it has been noted that not all flares are accompanied by CMEs and not all CMEs by flares \citep{Munro1979,Harrison1995,Yashiro2008,Wang2008}.

The CMEs from the Sun are known to create disturbances in the heliosphere, starting from their birth-place in the corona to the several AU distances away from the Sun \citep{Wang2000,Richardson2010}. They are found to interact with the atmosphere and magnetosphere of planets leading to severe space weather activity \citep{Wang2003b,Schwenn2006,Lundin2007,Echer2008,Baker2009,Mishra2015}. It is found that stellar mass loss also has a significant impact on stellar evolution, planetary evolution by increasing atmospheric erosion, the flux of cosmic rays incident on a planet's atmosphere and also on the larger-scale evolution of gas and dust in galaxies \citep{Willson2000,Lammer2007,Lammer2012,See2014}. So far it has not been possible to study the occurrence rate of CMEs from any star, except the Sun, purely based on the observations. The solar wind and CMEs from the Sun have been widely observed both remotely and in situ since the beginning of the space era. Thus, in the absence of detailed evidence of stellar CMEs, researchers of the stellar community often look to the Sun and gain insight into episodic stellar CMEs on magnetically active stars \citep{Aarnio2012,Osten2015,Cranmer2017}. But the question as to how the knowledge obtained from the solar observations may be applied to other stars is highly challenging.

The relationship between X-ray flare flux and corresponding CME mass has been examined by making a temporal and spatial correlation between them \citep{Andrews2003,Yashiro2005,Yashiro2006,Aarnio2011}. It was found that solar flares with higher energy have a higher probability to be CME-associated, and these CMEs are more massive.  The occurrence rate of CMEs varies with the solar magnetic cycle, which is the nearly 11-year periodic process that takes the Sun through subsequent periods of high (maximum) and low (minimum) activity. The relationship between the occurrence rate of CMEs and different proxies for solar magnetic variability may be different for different solar cycle and/or different phases of the cycles.

A commonly used proxy of solar activity is the number of sunspots and its latitudinal distribution, which is different for different cycles, and the distribution migrates towards the equator as the cycle progresses \citep{Solanki2000,Solanki2008,Hathaway2011}. The increased magnetic activity at higher latitudes and CMEs launched from there would facilitate mass loss from a star with a reduction in the spin down rate of the star \citep{Garraffo2015}. It would also be interesting to examine the latitudinal dependence of mass loss rate due to the divergence of CME latitude with the progression of the solar cycle. It would be interesting to establish a relation between the mass loss rate from the Sun and other proxies of solar magnetic variability such as Solar X-ray background luminosity which represents the state of solar corona \citep{Aschwanden1994}. Such studies may help to establish a proper scaling up factor to solar observations of magnetic variability, active regions, and coronal structures for understanding the mass-loss process on other stars having sun like corona.

The relative contribution of stellar CMEs into the stellar wind could be high for active stars \citep{Aarnio2012}, and quantification of this is important to understand the global mass-loss from the stars. For the case of the Sun, several attempts have been made to quantify the contribution of solar CMEs to background solar wind mass flux in the ecliptic at 1 AU \citep{Hildner1977,Howard1985,Jackson1993,Webb1994,Lamy2017}. In these studies the fractional contribution of CMEs to the solar wind was found to be different and it ranged from 5 to 16\% at the maximum of the solar cycles. The conflicting results on the fractional contribution of CMEs may be partly due to different chosen samples, instrument-dependent effects, and assumptions about the geometry of CMEs and calculation of CME mass flux. It would be interesting to reexamine the solar cycle variation of the fractional contribution of CMEs to solar wind mass flux in the ecliptic, and further extend the study to higher latitudes in the heliosphere.

\begin{figure*}
	\centering
		\includegraphics[scale=0.30]{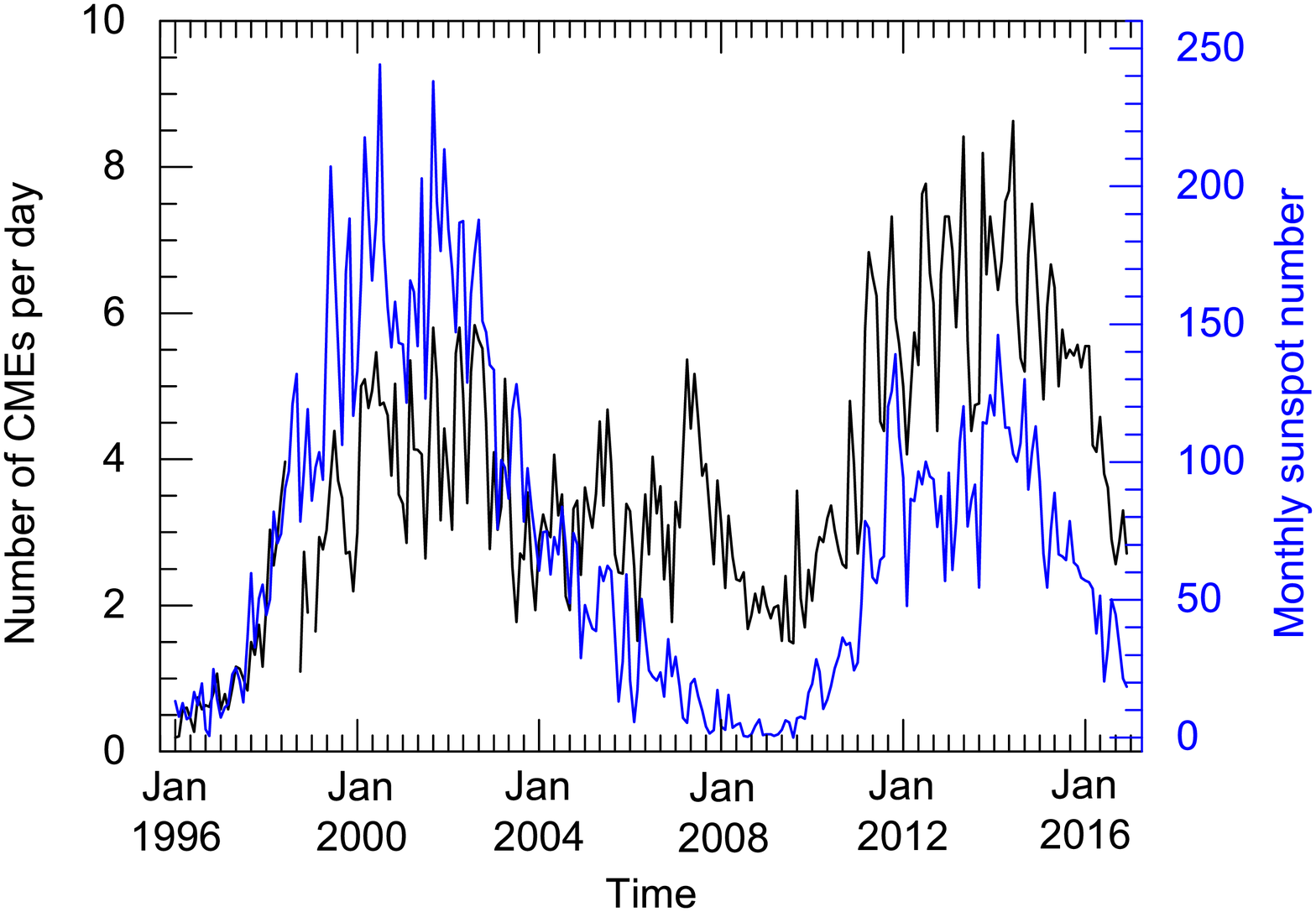}
		\hspace{5mm}
		\includegraphics[scale=0.30]{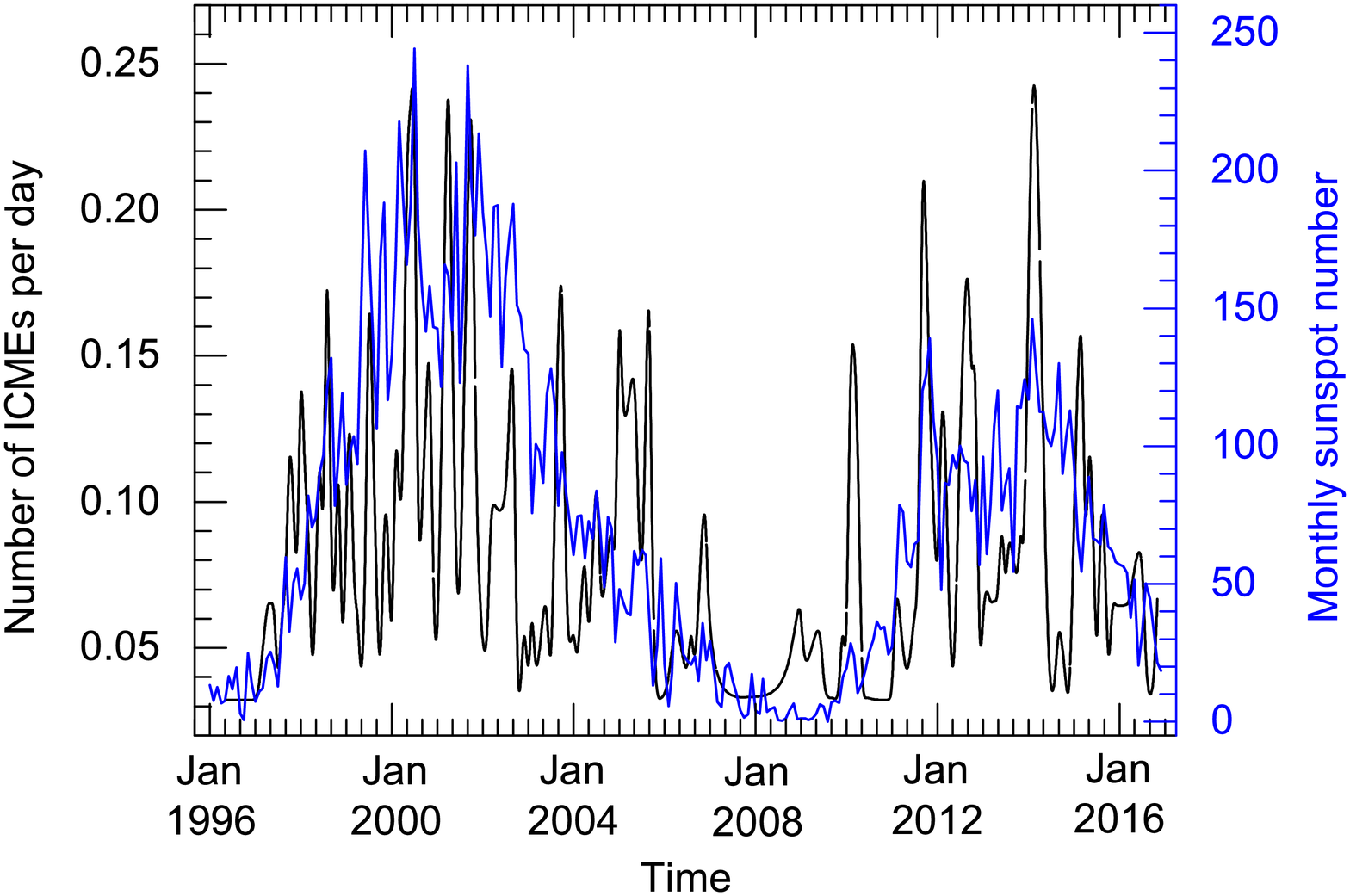}
		\caption{Left panel: the variation of rate of CMEs (on the left Y-axis in black) and monthly sunspot number (on the right Y-axis in blue) with time (on X-axis) during solar cycle 23 to 24 is shown. Right panel: Similar to the left panel, but for ICMEs.}
	\label{CME_ss_SCs}
\end{figure*} 

In the present study, we focus to investigate the evolution of mass loss rate via solar wind and CMEs from the Sun separately over solar cycles 23 and 24. We use solar X-ray background luminosity and sunspot number as a proxy to measure solar magnetic variability and find its relation with mass loss rate. Our analysis for understanding the mass loss via CMEs and solar wind is described in Section~\ref{cme_ml_ssxr} and 
~\ref{sw_ml_ssxr}, respectively. The result obtained from a new perspective on CME activity is compared with earlier studies of \citet{Luhmann2011}, \citet{Wang2014a}, \citet{Gopalswamy2015}, and \citet{Petrie2015}. We also reexamine the contribution of CMEs to the solar wind mass flux at 1 AU in the ecliptic and beyond over the solar cycles 23 and 24 using the method described in \citet{Webb1994} so that we can directly compare our results with earlier studies of \citet{Hildner1977},\citet{Howard1985}, \citet{Jackson1993}, \citet{Webb1994}, and \citet{Lamy2017}. The results and discussion of the study are presented in Section~\ref{resdis} and the conclusions are given in Section~\ref{conclu}.

\section{Variability in Occurrence of CMEs and Resulted Mass Loss Rate During Solar Cycle 23 and 24}
\label{cme_ml_ssxr}
Solar activity cycle has an average time period of 11 years and the most direct, key and commonly used indicator of solar activity is the number of sunspots on the solar photosphere. The reliable sunspot record exists for more than a century. Following Wolf's numbering scheme, the interval of the year 1755 to 1766 is traditionally numbered "1" solar cycle \citep{Wolf1861}. Solar cycle 23 is considered to span from August 1996 to December 2008 with its maximum around 2002 \citep{Joselyn1997,Detoma2000,Temmer2006}. It also had an extremely quiet and long solar minimum \citep{Janardhan2011,Bisoi2014}. The solar cycle 24 began in December 2008 reached the maximum around mid of 2014 and is currently in its declining phase \citep{Pesnell2008,Clette2014}. The solar cycle 24 is noted to be weaker than its preceding solar cycle in terms of disturbances in the convection zone, solar surface and the heliosphere \citep{Ramesh2010,Antia2010,Janardhan2011,Richardson2013,Gopalswamy2015}.

\subsection{Occurrence Rate of CME and ICME with Sunspot Number}
\label{cme_or_ss}
We obtained the data of CMEs observed during cycle 23 and 24 from the  Coordinated Data Analysis Workshops (CDAW) CME catalog (\url{https://cdaw.gsfc.nasa.gov/CME_list/}: \citealp{Yashiro2004}). The catalog lists the CMEs observed by Large Angle and Spectrometric Coronagraph (LASCO; \citealt{Brueckner1995}) on board the \textit{Solar and Heliospheric Observatory} (SOHO) mission. The LASCO initially carried three coronagraphs with overlapping fields of view (C1: 1.1-3 R$_\odot$, C2: 2-6 R$_\odot$, and C3: 3.7-32 R$_\odot$), among which the C1 could not survive after the temporary loss of the SOHO spacecraft in 1998 and therefore CMEs are identified using images from the C2 and/or C3 coronagraphs. In the catalog, the CME observations are not listed for the complete year of 2017 and beyond, therefore we limit our analysis for solar cycle 24 only for the period of the year 2009 to the year 2016. The sunspot data are obtained from the SIDC website 
(\url{http://sidc.oma.be/silso/datafiles}: \citealp{Clette2014}). To calculate the occurrence rates of CMEs, we include all the CME events listed in the catalog regardless of their morphological and kinematic characteristics, even if they are classified as "very poor" events. During January 1996 to December 2016, the CDAW catalog lists 28315 CMEs observed with SOHO/LASCO.

We counted the number of CMEs in a calender month and then estimated the average number of CMEs per day. The occurrence rate of CMEs, as well as monthly sunspot number for solar cycle 23 and 24, is shown in the left panel of Figure~\ref{CME_ss_SCs}. For the solar cycle 23, we note that the variation in the CME rate follows the sunspot number till the year 2003. After this, although the sunspot cycle 23 continued to decline up to the end of the year 2008, the CME rate did not decline. Surprisingly, the CME rate shows a prominent peak at the beginning of the year 2007 and then shows a modest decrease for only one year. After entering the solar cycle 24, the variation in the CME occurrence rate follows the number of sunspots closely. The sunspot cycle 24 has a significantly smaller amplitude than the corresponding phase of the previous sunspot cycle 23. But, CME occurrence rate during cycle 24 is roughly equal to or sometimes slightly higher than that during the corresponding phase of cycle 23. It is worth mentioning that the rate of CMEs for the period of 2012-2014 is increased by a factor of $\approx$1.6 while the sunspot number is decreased by a factor of $\approx$1.7 than that in 2000-2002 years.

It is to be noted that we used the data from the CDAW CME catalog which is based on the visual (i.e, by the human eye) identification of CMEs from the white light coronagraphic images of LASCO C2 and C3 by the designated observers. Such manual identification and compilation of CMEs are subjective and observer-dependent. Even for a single observer, the visual identification ability may change over time due to continued work-experience of the observer. Therefore, in addition to the CDAW catalog, several attempts have been made to automatically identify the CMEs in near real-time data. The widely used automated CME catalogs are Computer Aided CME Tracking (CACTus) from the images of both LASCO C2 and C3 \citep{Robbrecht2004}, Automatic Recognition of Transient Events and Marseille Inventory from Synoptic maps (ARTEMIS) from the images of LASCO C2 only \citep{Boursier2005} and Solar Eruptive Event Detection System (SEEDS) from the images of LASCO C2 only \citep{Olmedo2008}. Thus, the catalogs using LASCO observations can be classified as C2+C3 catalogs (e.g., manual CDAW and automated CACTus) and C2-only catalogs (e.g., automated ARTEMIS and SEEDS). The correlation of CME rate with sunspot number during solar cycle 23 to 24 have been discussed earlier by \citet{Wang2014a} using only SEEDS catalog, by \citet{Lamy2014} using only ARTEMIS catalog, by \citet{Gopalswamy2015} using only CDAW catalog, and by \citet{Petrie2015} using three catalogs as CDAW, ARTEMIS, and SEEDS.

Using the SEEDS C2-only catalog, \citet{Wang2014a} shows a steep increase in the occurrence rate of CMEs from 2010 onward, and the CME rate is almost equal in the maximum of solar cycle 23 and 24. They argue that the increased rate of CMEs in SEEDS is an artifact due to increased LASCO image cadence from 60 images per day to about 100 images per day in 2010 August. Recently, \citet{Hess2017} artificially lowered the cadence during 2010 to 2015 to keep the same cadence for the SEEDS algorithm during the entire LASCO observations and have shown that the CME rate is almost consistent with sunspot number for both the cycles 23 and 24. Further, it has been shown using the ARTEMIS C2-only catalog that CME rate has a high correlation with the sunspot number during both the cycles 23 and 24 \citep{Lamy2014}. However, \citet{Petrie2015} concluded a real increase in CME rate around 2003 using CDAW C2+C3 and CACTus C2+C3 catalogs while the increased CME rate appeared around 2010 using SEEDS C2-only catalog. They showed that the pattern of increased detection around 2003 persists even if the small and faint ejections which are likely to be observer-dependent are excluded from the CDAW catalog after the middle of cycle 23. Further, using CDAW C2+C3 catalog, the study of \citet{Gopalswamy2015} on the halo CMEs, a sub-set of CMEs for which image cadence is not an issue, shows an increased rate of CMEs in cycle 24. Recently, using the data from automated CACTus C2+C3 catalog, it has been noted that peak in the occurrence rate of CME during the maximum of solar cycles 23 and 24 is the same, but the peak in solar cycle 24 is more extended in time than that in cycle 23 \citep{Compagnino2017}. We extend previous studies to reexamine whether or not the CME occurrence rate has really increased after the middle of solar cycle 23. For this, we adopt a  physically more meaningful approach of inferring the CME activity in terms of mass loss rate via CMEs instead of CME number. The details of our approach given in Section~\ref{cme_ml_ss} and 
~\ref{cme_ml_sxr} shows a true physical increase in CME activity which is in agreement to \citet{Petrie2015}.

The CMEs launched from the Sun are termed as ICMEs when they propagate in interplanetary medium \citep{Dryer1994,Zurbuchen2006,Crooker2006}. ICMEs are identified in in situ observations based on the properties of plasma, magnetic field and compositional parameters \citep{Zurbuchen2006}. To better understand the rate of CMEs from the Sun, we estimated the number of ICMEs per day in the ecliptic plane at 1 AU using the ICME catalog compiled by Ian Richardson and Hilary Cane \citep{Cane2003,Richardson2010}. From the catalog, we counted the number of ICMEs observed during each calendar month and then calculated the average number of ICMEs per day. The occurrence rate of ICMEs for solar cycle 23 and 24 is shown in the right panel of Figure~\ref{CME_ss_SCs}. We note that ICME rate follows the sunspot cycle 23, however, the rate is relatively higher in the declining phase of solar cycle 23 during the year 2005 to 2007. We note that ICME rate is smaller during the rising phase of the solar cycle 24 than the corresponding phase of cycle 23. The total count of ICMEs during the first 7 years of cycle 24 is decreased by 40\% of that during the same interval of the previous cycle. However, during the maximum of cycle 24, the rate of ICME reaches up to the same value as during the maximum of cycle 23. Further, we find that the near-Sun average speed of the CMEs during solar cycle 23 is significantly higher than that during the corresponding phase of solar cycle 24. The average speed of the ICMEs during solar cycle 24 is only slightly lower than that during solar cycle 23. 

\subsection{Mass Loss Rate via CMEs Versus Sunspot Numbers}
\label{cme_ml_ss}
There is no general consensus on the criteria in identifying a CME from the Sun. Therefore, under-/overcount of the number of CME using white-light coronagraphic observations is quite observer/algorithm dependent \citep{Yashiro2004}. The manual CME catalog may miss several narrow and slow CMEs due to human error while the automated catalog may falsely include streamer deflections, internal parts of a CME, and gusty flows following a CME as separate events \citep{Robbrecht2009a}. Therefore, estimating the occurrence rate of CME is very subjective as the manual and automated catalogs differ in CME statistics and characteristics. Also, the identification of ICMEs in in situ observations is not objective as it is based on the magnetic field, dynamic and composition signatures in solar wind plasma \citep{Richardson2010}. The rate of ICMEs in the ecliptic plane may be dependent on the solar cycle variations in the CME source locations at the Sun \citep{Gopalswamy2003,Zhao2003}. Thus, CME and/or ICME occurrence rate may not truly represent the CME activity over different phases of the solar cycles. However, it has been shown that manual CDAW catalog and automated CACTus catalog have a good agreement on characteristics and occurrence rate of wide CMEs observed in the LASCO field of view \citep{Yashiro2008a,Robbrecht2009a}. The narrow faint ejections are not much massive and they contribute only a little to mass loss from the Sun via CMEs. Therefore, it would be physically more meaningful to measure solar activity in terms of the total mass loss from the Sun due to CMEs instead of CME occurrence rate. In this case, the bias of observer to exclude or include very faint CMEs as poor events would be negligible on the total mass loss. It is also noted that a majority of CMEs are associated with sunspot regions which drift from higher latitude to lower latitude as the cycle progresses \citep{Harrison1990,Subramanian2001}. We therefore examine the variation in CME mass loss rate with solar latitude and its changes over the solar cycle. It is expected that the comparison of CME mass loss rate between solar cycle 23 and 24 would establish the physical difference in CME activity over the cycles.

\subsubsection{Latitude and Mass of the CMEs}
The CMEs are observed in the plane of the sky from \textit{SOHO}/LASCO,  therefore all apparent spatial parameters of the CMEs listed in CDAW catalog are a projection of the real values onto that plane. We used the central position angle of the CME, as listed in the catalog, to determine its apparent latitude as $\delta$ = 90$^\circ$ - CPA for 0$^\circ$ $\le$ CPA $<$ 180$^\circ$, and $\delta$ = CPA - 270$^\circ$ when 180$^\circ$ $\le$ CPA $<$ 360$^\circ$ \citep{Yashiro2004}. In the catalog, the full halo CMEs do not have a CPA value which poses a difficulty in determining their apparent latitude. Further, the mass estimates of full halo CMEs are very uncertain. Therefore, we excluded such full halo CMEs from our statistics and attempt to examine the variation in mass loss rate via CMEs from different latitudes with sunspot numbers. Since full halo CMEs constitute less than 4\% of all CMEs \citep{Gopalswamy2004}, it is expected that they will have not much effect on our statistical results. We classified the CMEs into three groups, (i) CMEs having latitude between -30$^\circ$ to 30$^\circ$ (ii) CMEs having latitude between -60$^\circ$ to 60$^\circ$ and (iii) CMEs having latitudes between -90$^\circ$ to 90$^\circ$.

\begin{landscape}
\begin{figure}  
	\centering
		\includegraphics[scale=0.26]{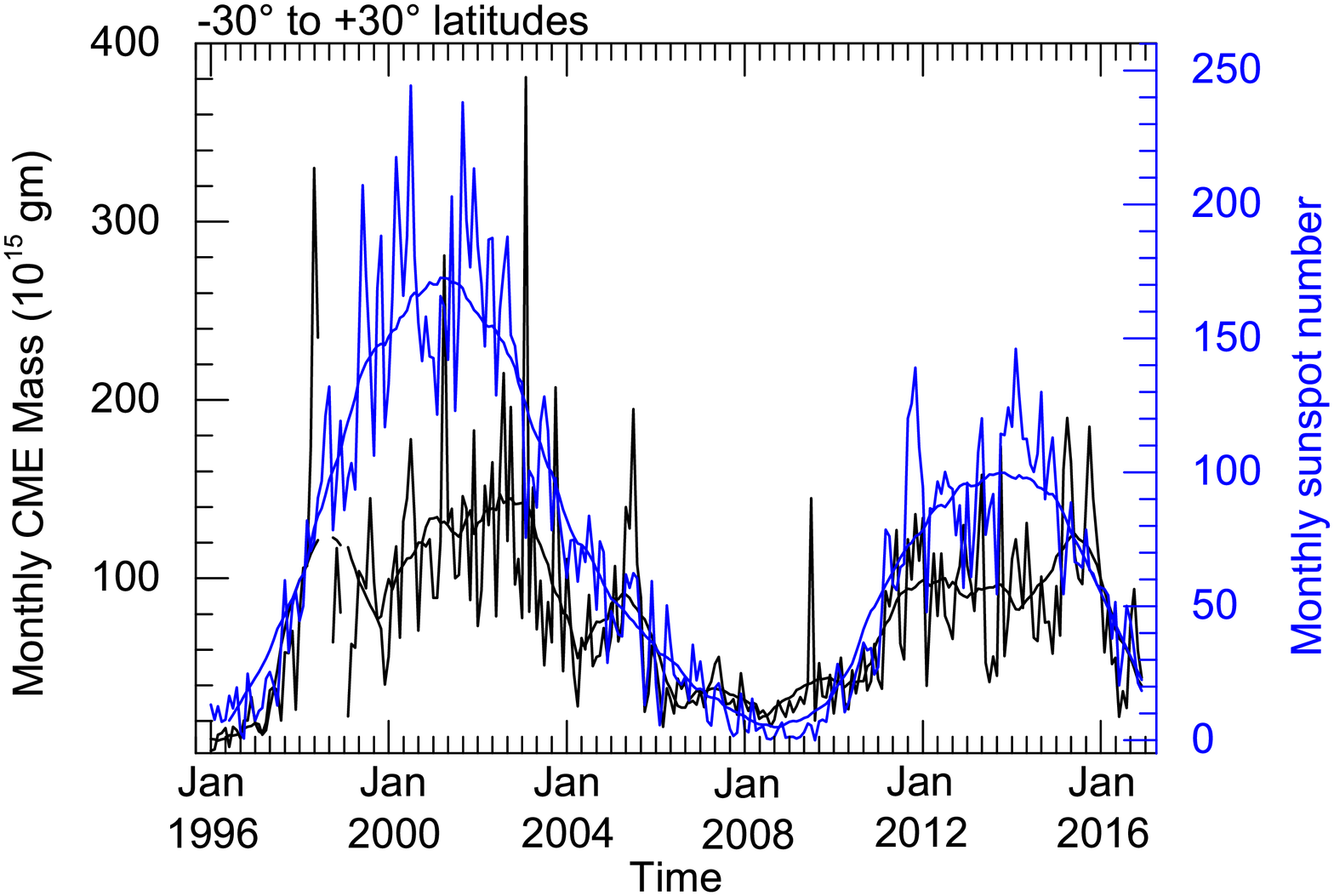}
		\hspace{4mm}
		\includegraphics[scale=0.26]{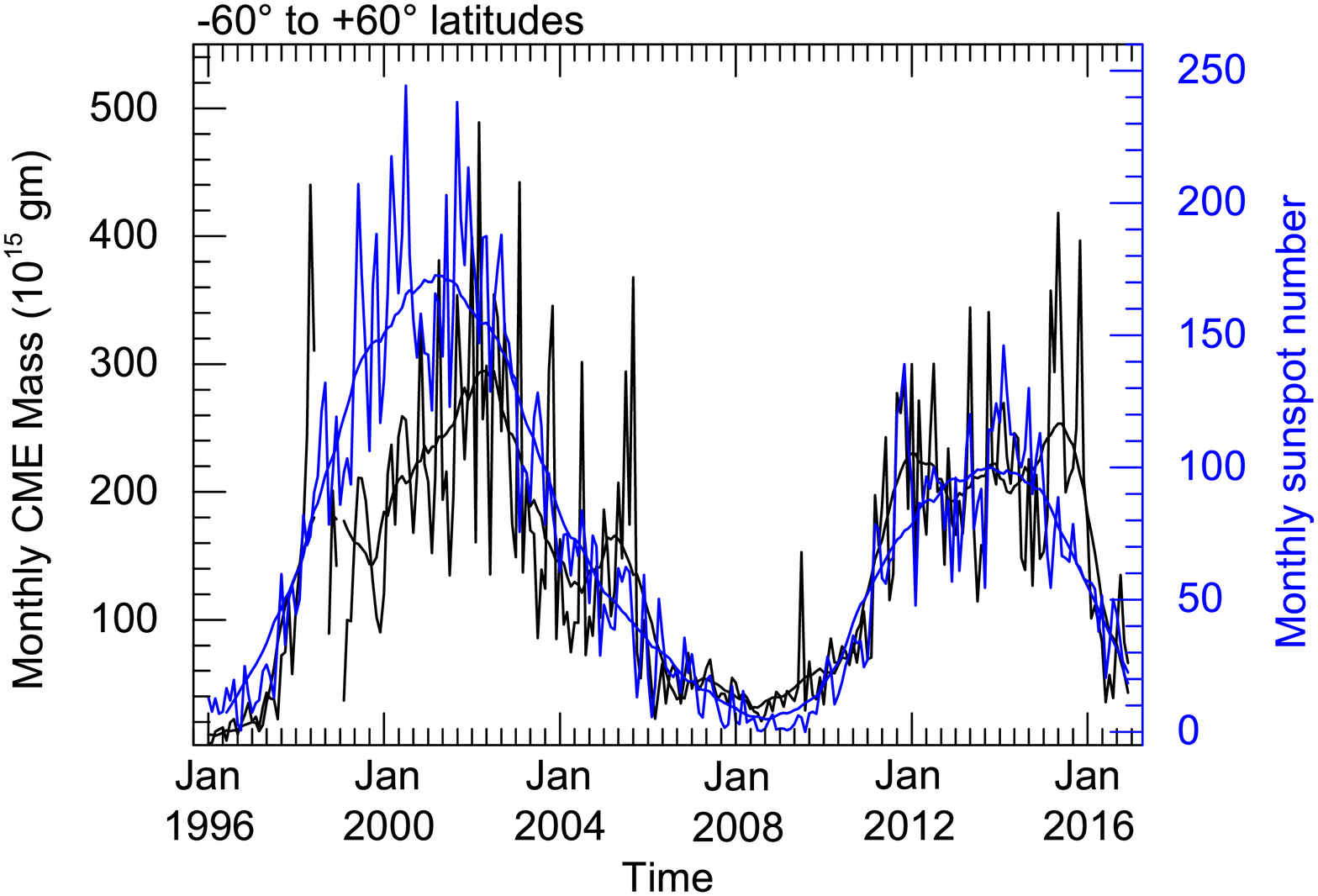}
		\hspace{4mm}
		\includegraphics[scale=0.26]{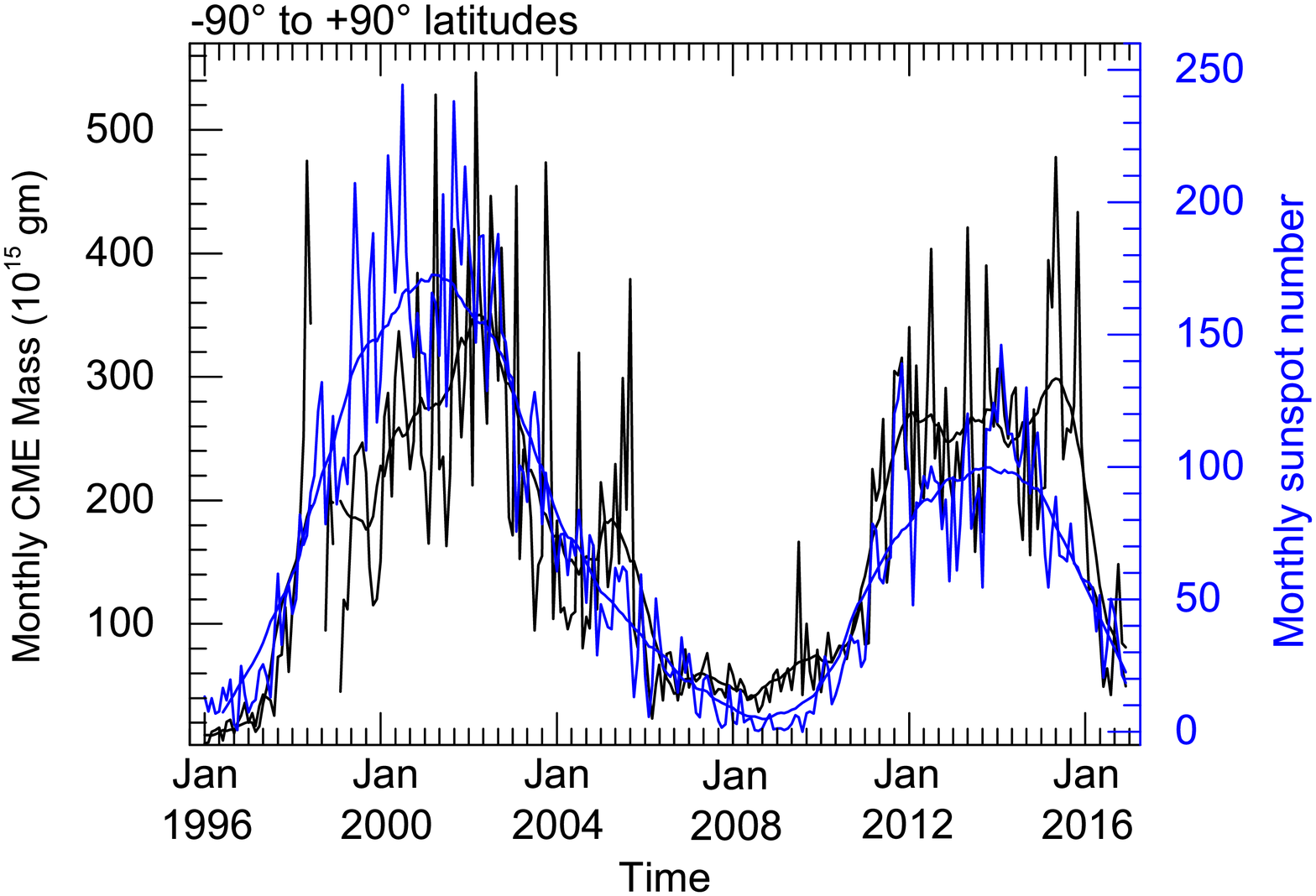}\\
		\vspace{3mm}
		\includegraphics[scale=0.26]{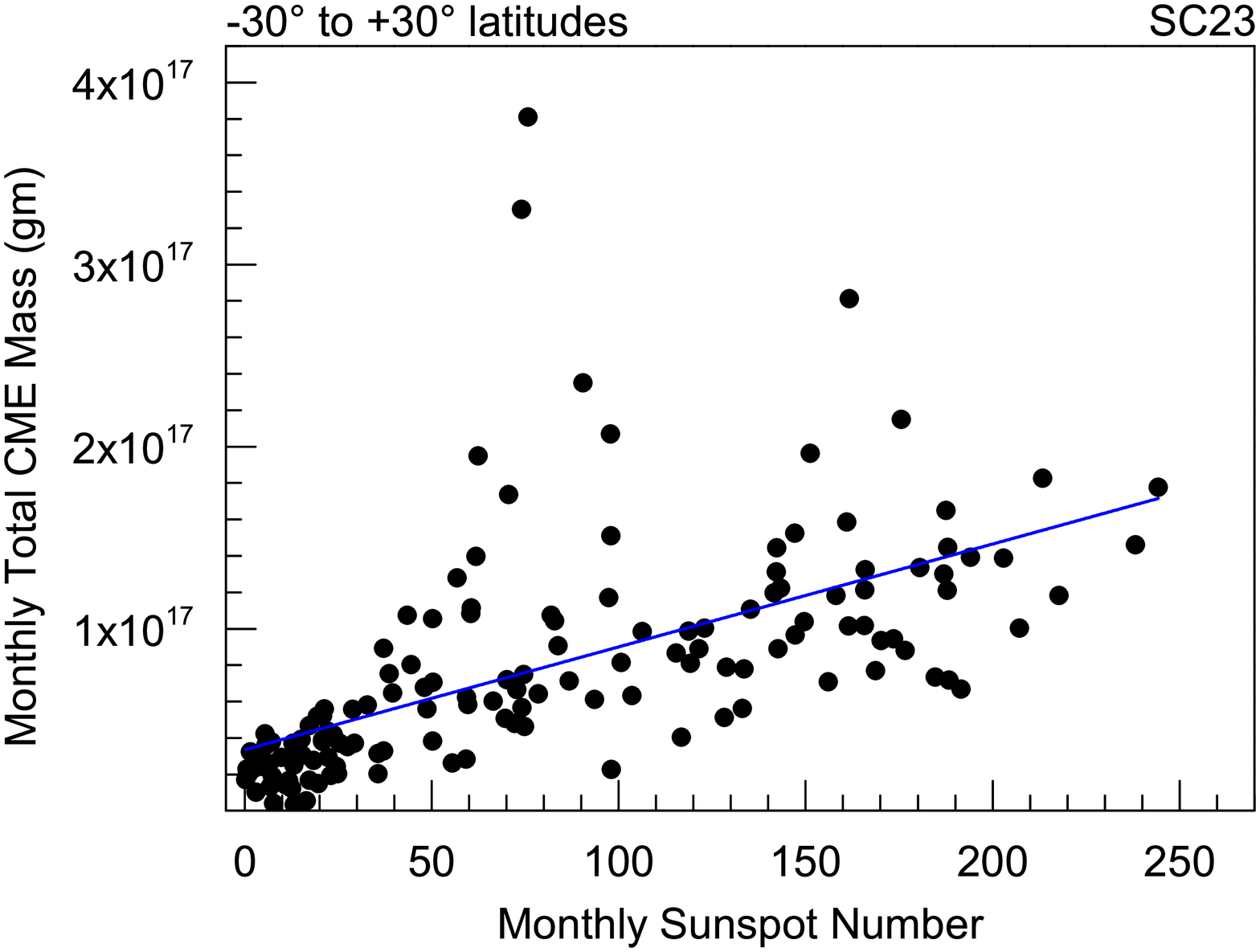}
		\hspace{7mm}
		\includegraphics[scale=0.26]{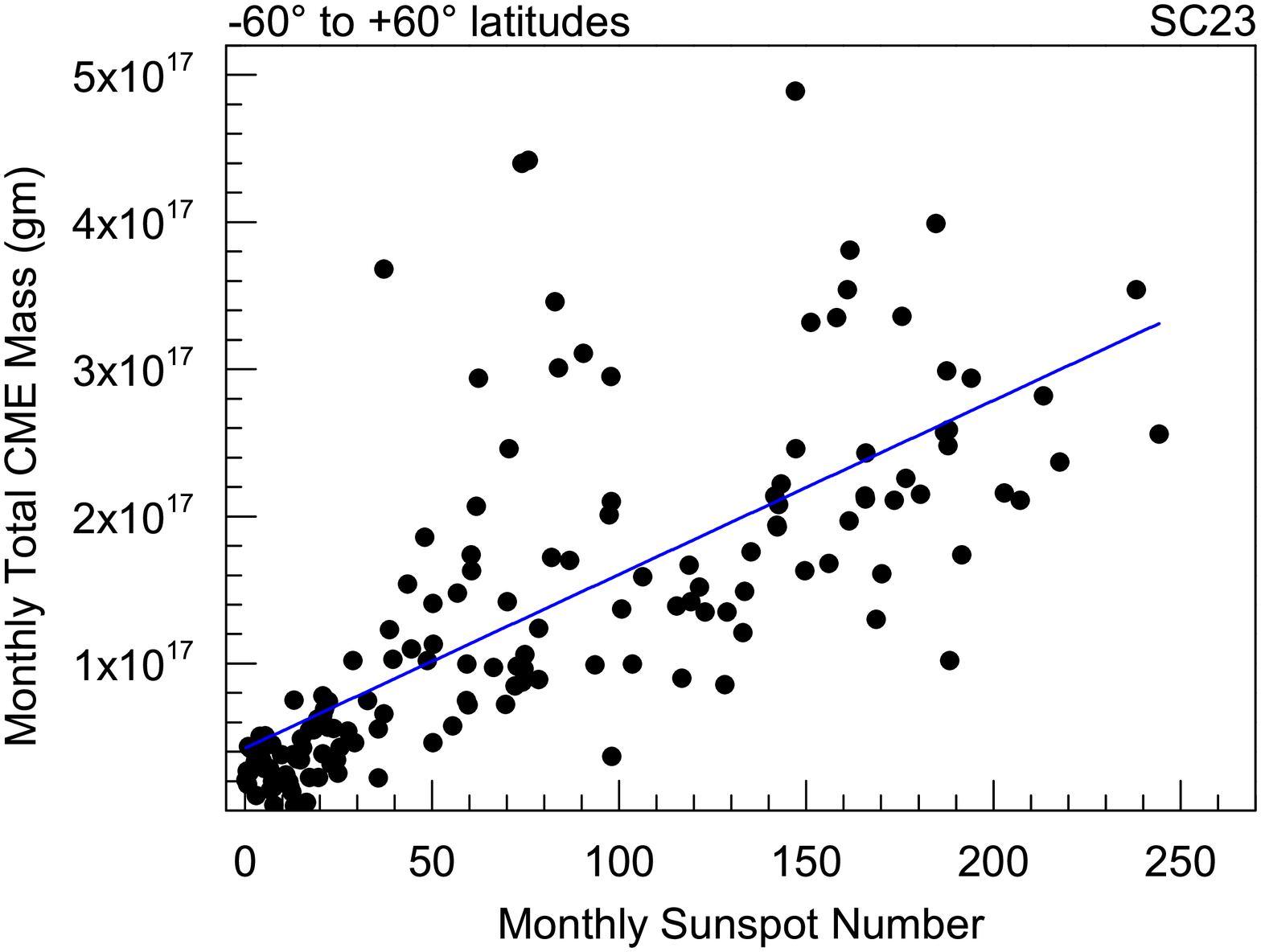}
		\hspace{7mm}
		\includegraphics[scale=0.26]{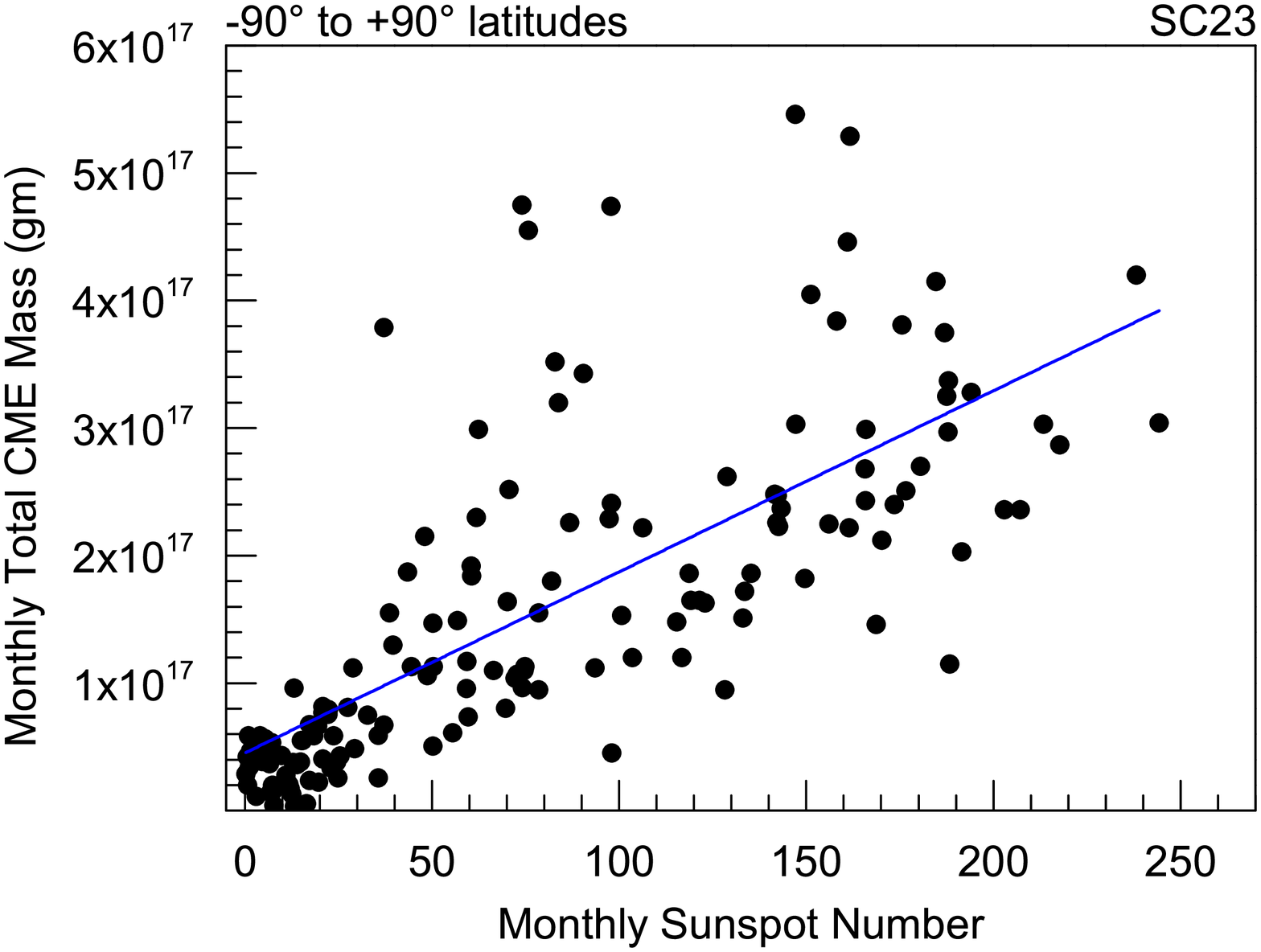}  \\
		\vspace{3mm}
		\includegraphics[scale=0.26]{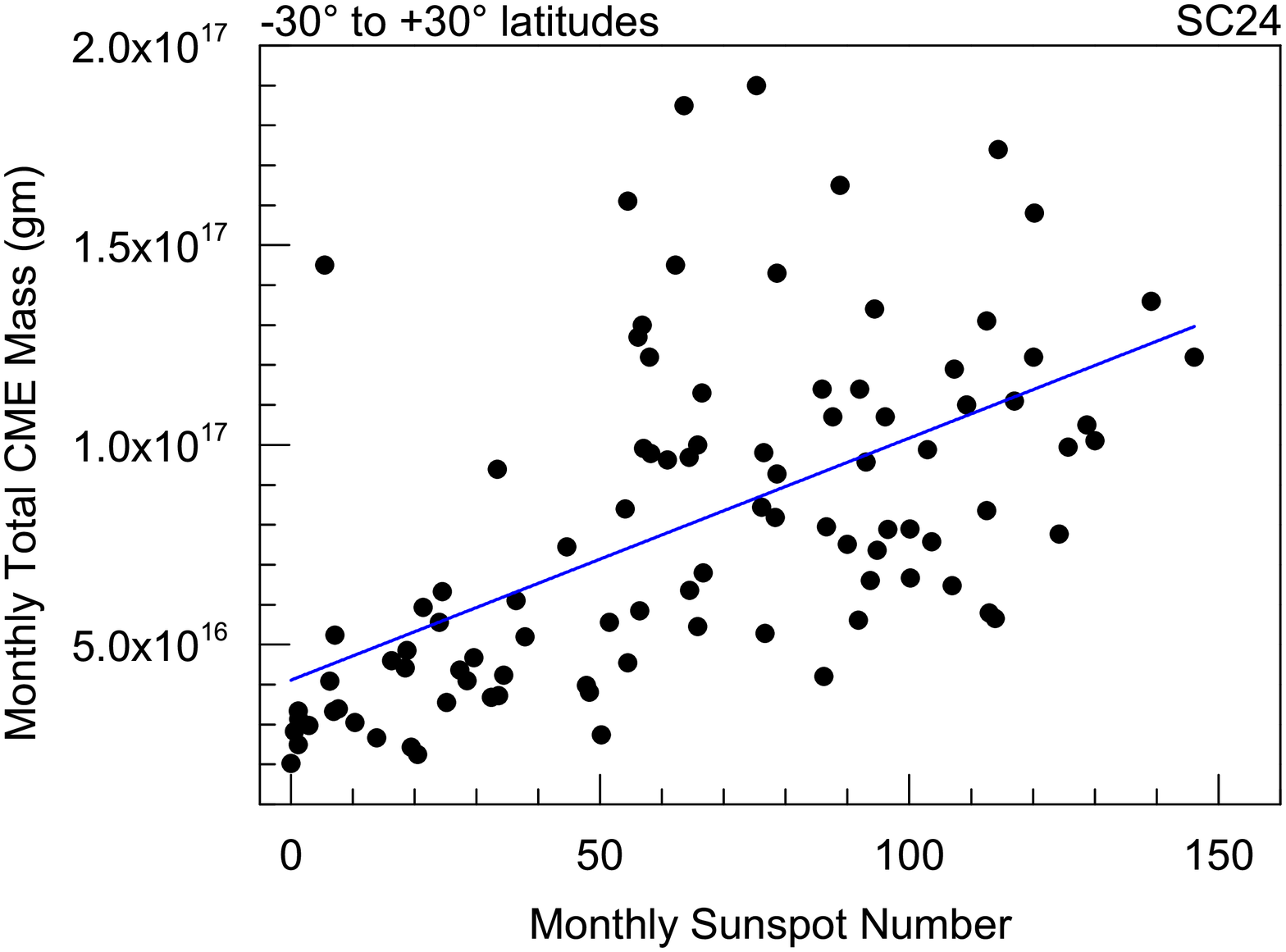}
		\hspace{7mm}
		\includegraphics[scale=0.26]{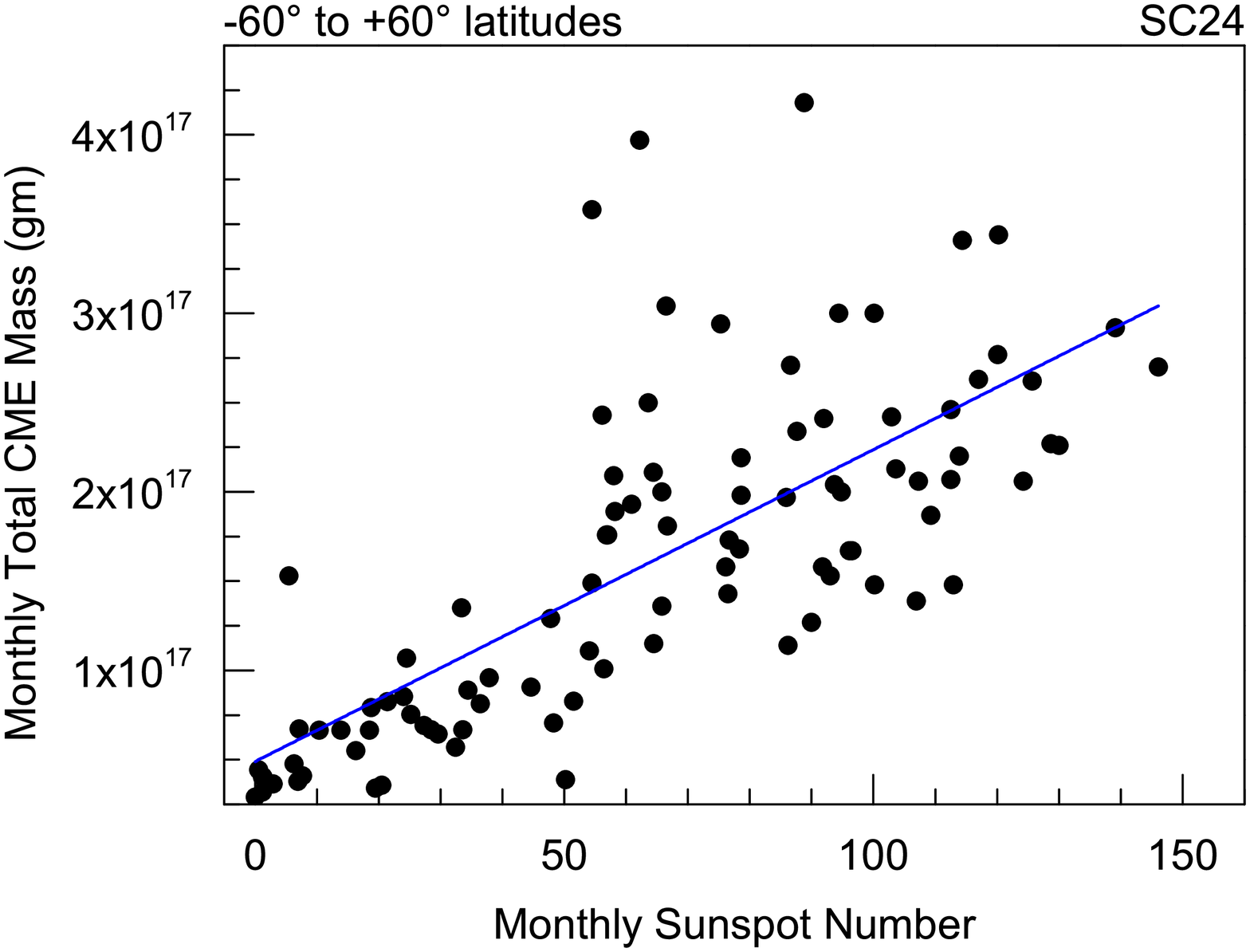}
		\hspace{7mm}
		\includegraphics[scale=0.26]{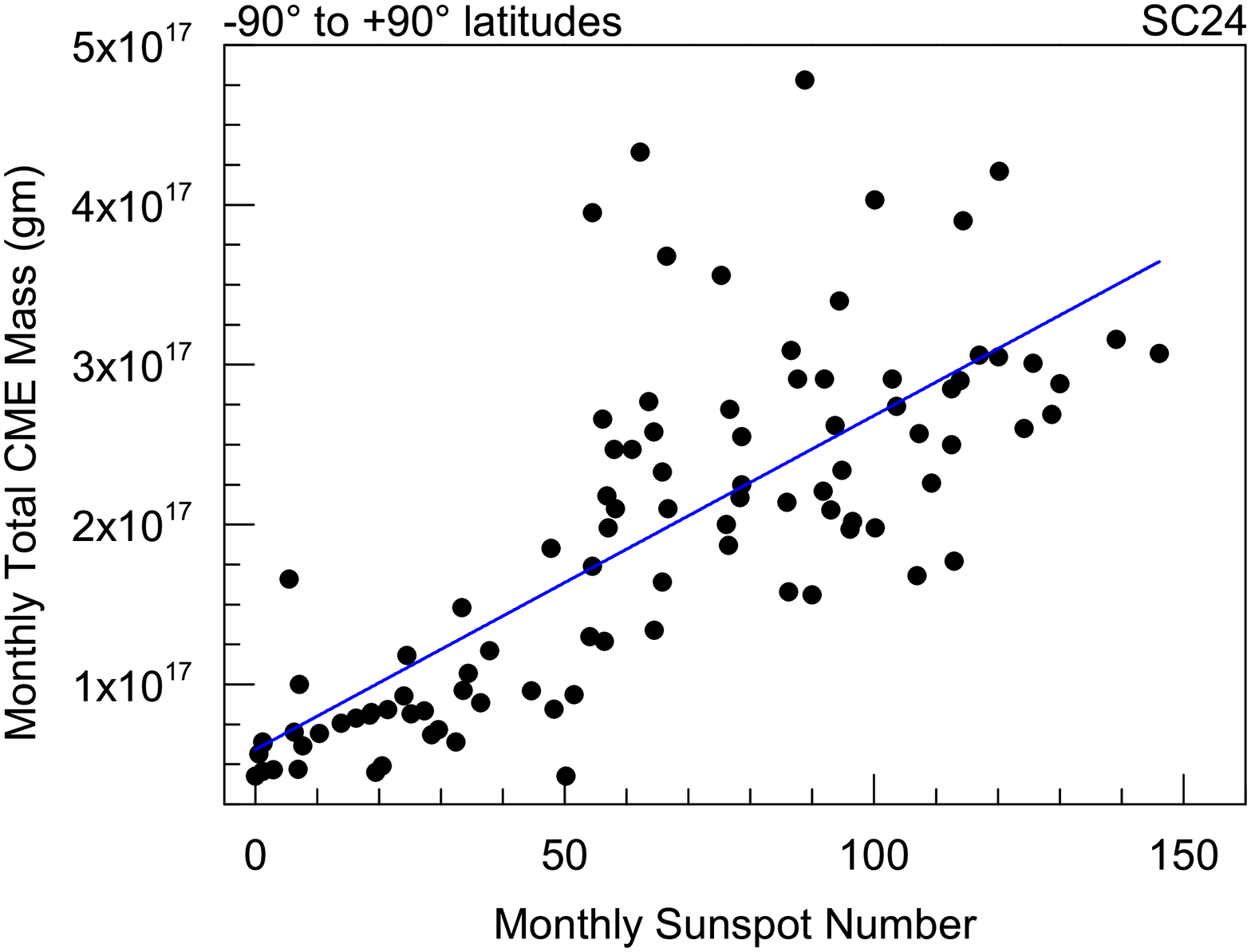} \\
		\caption{Left panel: the variation of rate of mass loss (on the left Y-axis in black) due to CMEs from latitude within 
		-30$^\circ$ to 30$^\circ$ and monthly sunspot number (on the right Y-axis in blue) with time (on X-axis) during 1996     to 2016 (i.e, solar cycle 23 to 24) is shown in the top panel. The smoothed values of actual measurements 
		are also overplotted (top panel). Scatter plot between monthly sunspot number and monthly mass due to 
		CMEs from latitude within -30$^\circ$ to 30$^\circ$ with a fitted regression line (in blue) for solar cycle 23 
		(middle panel, labeled as SC23) and solar cycle 24 (bottom panel, labeled as SC24) is shown.    
		Central panel: Similar to the left panel, but the rate of mass loss due to CMEs from latitude within -60$^\circ$ 
		to 60$^\circ$ is plotted. 
		Right panel: Similar to the left panel, but the rate of mass loss due to CMEs from latitude within 
		-90$^\circ$ to 90$^\circ$ (i.e, anywhere from the Sun) is plotted. }
	\label{CME_massloss_ss_SC2324}
\end{figure}
\end{landscape}

During January 1996 to December 2008, CDAW catalog lists 14018 CMEs which includes "very poor" events also. The CMEs classified as very poor are extremely small, faint, and appear only in a few image frames. The mass is determined by subtracting a pre-event background image from the image containing a CME, and the mass associated with each pixel is derived from the excess brightness using the theory of Thomson scattering \citep{Minnaert1930,Howard2009,Howard2012}. Then the excess mass at each pixel is summed over the angular width of the CME. The estimated mass is underestimated as the CME is assumed to be located in the sky plane \citep{Vourlidas2000,Vourlidas2010}. The catalog provides the mass estimates for about 50\% of the total number of CMEs. Based on the number of CMEs during a year and their mass estimates, we determined the yearly averaged mass of a CME. We assign this yearly averaged mass to CMEs of that year for which mass estimates are not listed in the catalog. Thus, for all the CMEs in our sample, we have their representative source region latitude and mass estimates.

\subsubsection{CME Mass Loss Rate and Sunspot Cycles 23 and 24} 
\label{cme_ml_ss_main}
We calculated the total mass ejected from the Sun during a calendar month due to CMEs from the apparent latitude within -30$^\circ$ to 30$^\circ$, -60$^\circ$ to 60$^\circ$ and -90$^\circ$ to 90$^\circ$. The variation in mass loss rate due to these three groups of the CMEs along the progression of solar cycle 23 and 24 is shown in the top panel (left, central and right panels) of the Figure~\ref{CME_massloss_ss_SC2324}. From the top panels of the figure, we note that monthly mass loss due to CMEs from latitudes within -30$^\circ$ to 30$^\circ$ is significantly lower relative to sunspot number for the period of 1999 to 2002, i.e., during the rising and maximum phase of sunspot cycle 23. This lower mass loss rate becomes less obvious when the CMEs from increasingly higher latitudes within -60$^\circ$ to 60$^\circ$ and -90$^\circ$ to 90$^\circ$ are included. Similarly, a significantly lower mass loss rate during 2012-2014, i.e., the rise of sunspot cycle 24 is noted.

To quote, the maximum mass loss rate via CMEs from -30$^\circ$ to 30$^\circ$ at the peak of cycle 23 is 
1.5 $\times$ 10$^{17}$ gm month$^{-1}$ while it increases to 3.5 $\times$ 10$^{17}$ gm month$^{-1}$ when CMEs from all over the latitudes are included. We note that during the maximum of the cycle 23, CMEs originating from the latitudes within -30$^\circ$ to 30$^\circ$ and -60$^\circ$ to 60$^\circ$ contribute around 45\% and 90\% of the total CME mass loss, respectively. Also, the maximum mass loss rate via CMEs from -30$^\circ$ to 30$^\circ$ at the peak of the cycle 24 is 1.3 $\times$ 10$^{17}$ gm month$^{-1}$ while it increases to 2.7 $\times$ 10$^{17}$ gm month$^{-1}$ when CMEs from all over the latitudes are included. We note that during the maximum of cycle 24, the fractional contribution of CMEs from the latitudes within -30$^\circ$ to 30$^\circ$ and -60$^\circ$ to 60$^\circ$ to the total CME mass loss is around 40\% and 80\%, respectively. This suggests that the CMEs having apparent latitudes higher than 60$^\circ$ contribute to approximately 10\% and 20\% of the total mass loss rate in cycles 23 and 24, respectively, especially during the rise and maximum of the cycles. However, a significant fraction of mass loss via CMEs is from lower-mid latitudes.

We also noted that the mass loss rate per sunspot number is higher for solar cycle 24 than that in solar cycle 23 (top-right panel of Figure~\ref{CME_massloss_ss_SC2324}). The sunspot number at the maximum of cycle 24 is decreased by around 40\% compared to that during the maximum of previous cycle 23. However, the mass loss rate during the maximum of sunspot cycle 24 due to CMEs from latitudes within -30$^\circ$ to 30$^\circ$, -60$^\circ$ to 60$^\circ$, and -90$^\circ$ to 90$^\circ$ is decreased by only around 25\%, 20\%, and 15\%, respectively. This implies that the mass loss from higher latitudes is relatively larger during the maximum of cycle 24 than that during the previous maximum. Our finding suggests that a decrease in mass loss rate in solar cycle 24 is not as efficient as it is for the sunspot number. In general, the rate of mass loss due to CMEs is increased since 2003. A similar, but stronger, increase in occurrence rate of CMEs was noted since 2003 as shown in the left panel of Figure~\ref{CME_ss_SCs}. Therefore, it seems that the increased rate of CMEs consists of a good fraction of ejections with smaller mass.

To express the monthly mass loss rate ($dM_{CME}/{dt}$) in terms of monthly averaged sunspot number ($S$), we fit the observations with a mathematical expression given by,

\begin{equation}
 \frac{dM_{CME}}{dt}=5\times 10^{14} (c_{1} S + c_{2})    \text{~~~~gm month$^{-1}$}
	\label{eqcmlrss}
\end{equation}

where c$_{1}$ and c$_{2}$ are the constants. The fitted value of constants (c$_{1}$ and c$_{2}$), correlation coefficient ($r$) and coefficient of determination ($r^{2}$) obtained using Equation~\ref{eqcmlrss} for solar cycles 23 and 24 are shown in the top panel of Table~\ref{coef_func}. The value of $r$ measures the strength and direction of a linear relationship between two variables. The value $r^{2}$ is a statistical measure of how close the data are to the fitted regression line. The relationship between monthly mass loss and monthly sunspot number for cycle 23 and 24 is shown using scatter plots in the middle and bottom panels of Figure~\ref{CME_massloss_ss_SC2324}, respectively. The scatter plots for three groups of the CMEs from latitudes within -30$^\circ$ to 30$^\circ$, -60$^\circ$ to 60$^\circ$, and -90$^\circ$ to 90$^\circ$ with fitted regression line is shown in the left, central and right panels, respectively.

From Table~\ref{coef_func}, for both the cycle 23 and 24, we note that the obtained values of c$_{1}$, c$_{2}$, $r$ and $r^{2}$ increase when CMEs originating from the higher latitudinal range is included. For solar cycle 23, Equation~\ref{eqcmlrss} explains around 36\%, 51.8\%, and 54.8\% of the variability in data of monthly mass loss due to CMEs from latitudes within -30$^\circ$ to 30$^\circ$, -60$^\circ$ to 60$^\circ$, and -90$^\circ$ to 90$^\circ$, respectively. Similarly, for solar cycle 24, around 32.5\%, 54.8\%, and 57.8\%  of the variability in the observed mass loss due to CMEs within three groups of the increasingly higher latitudinal range are explained. Thus, the monthly averaged sunspot ($S$) number, even located at lower latitudes (generally within -30$^\circ$ to 30$^\circ$), is a good proxy for the CMEs from lower as well as higher latitudes. The values of constants obtained for cycle 24 are larger than that for the corresponding latitudinal range of cycle 23. This suggests a stronger dependence of mass loss rate via CMEs on the sunspot numbers in cycle 24 than the previous cycle. Based on the comparison of occurrence rate (left panel of Figure~\ref{CME_ss_SCs}) and total mass loss via CMEs during cycle 23 and 24 (top-right panel of Figure~\ref{CME_massloss_ss_SC2324}), it seems that relative contribution of CMEs, consisting of a large number of smaller ejections, from higher latitudes increases since 2003 and continues during weaker sunspot cycle 24. Thus, our finding is in agreement to an earlier study of \citet{Petrie2015} which suggested for a true physical increase in CME detection rate since the middle of solar cycle 23.

\begin{figure}
	\centering
		\includegraphics[scale=0.30]{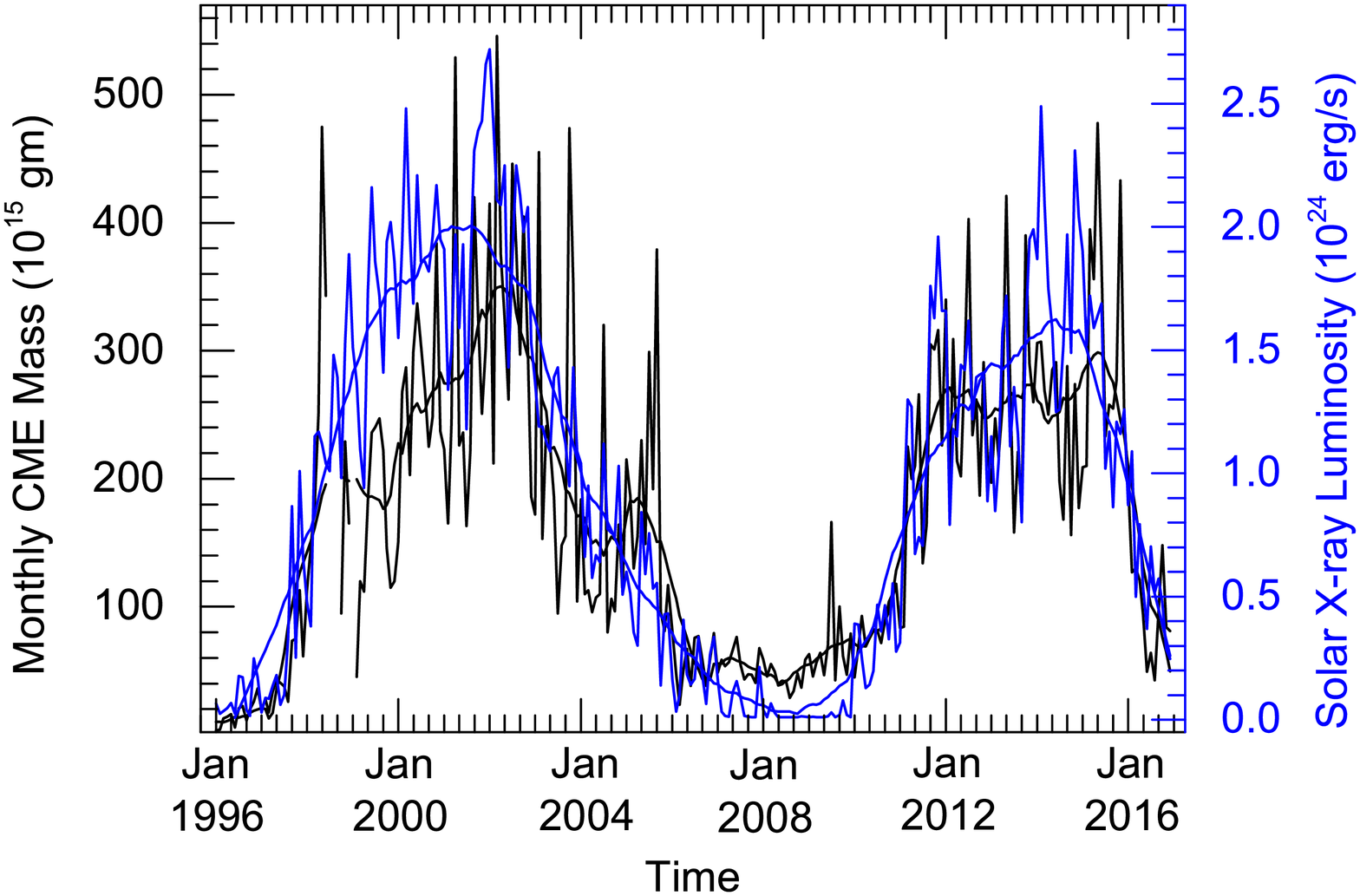} 
		\vspace{2mm}
		\includegraphics[scale=0.30]{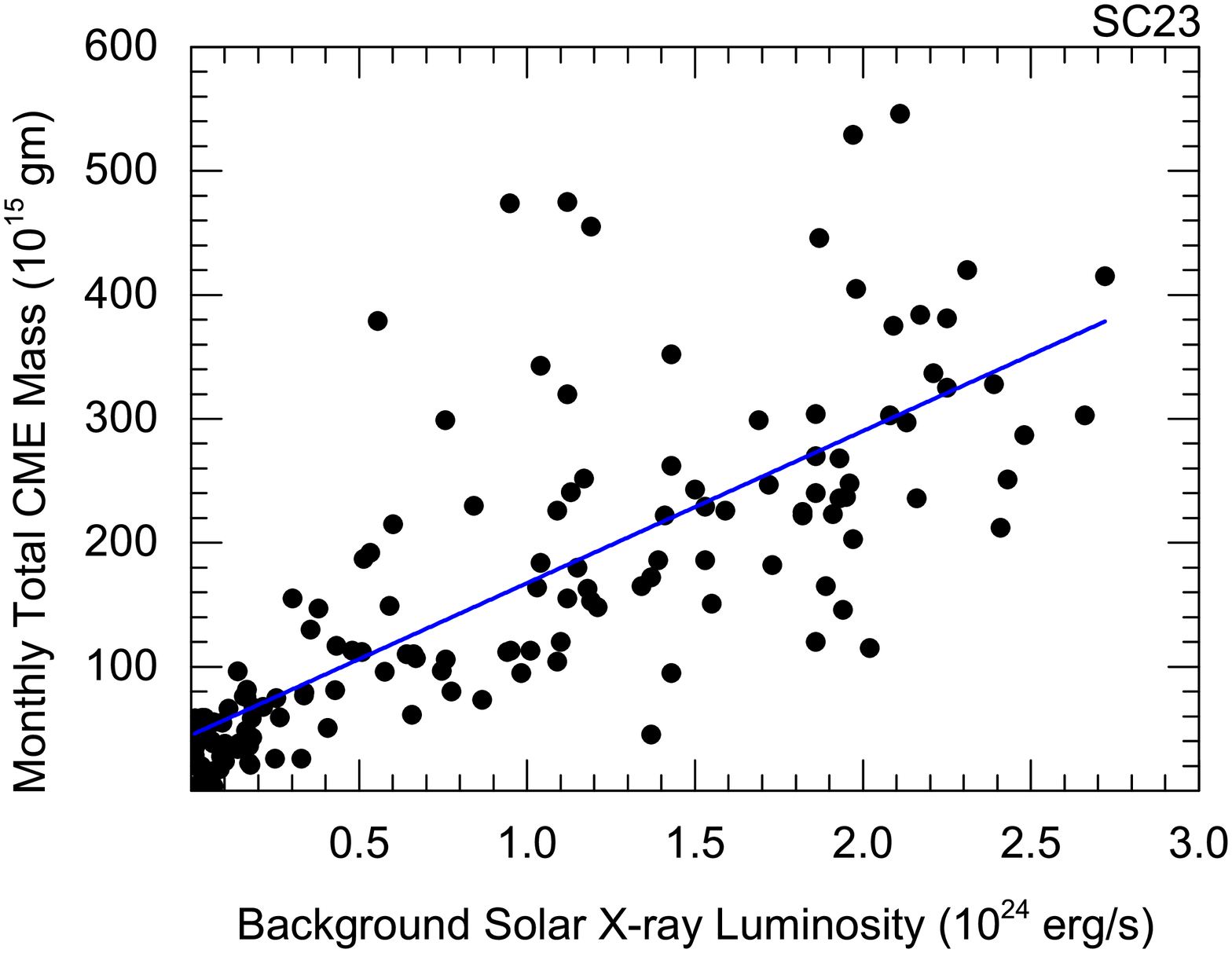}
		\vspace{2mm}
		\includegraphics[scale=0.30]{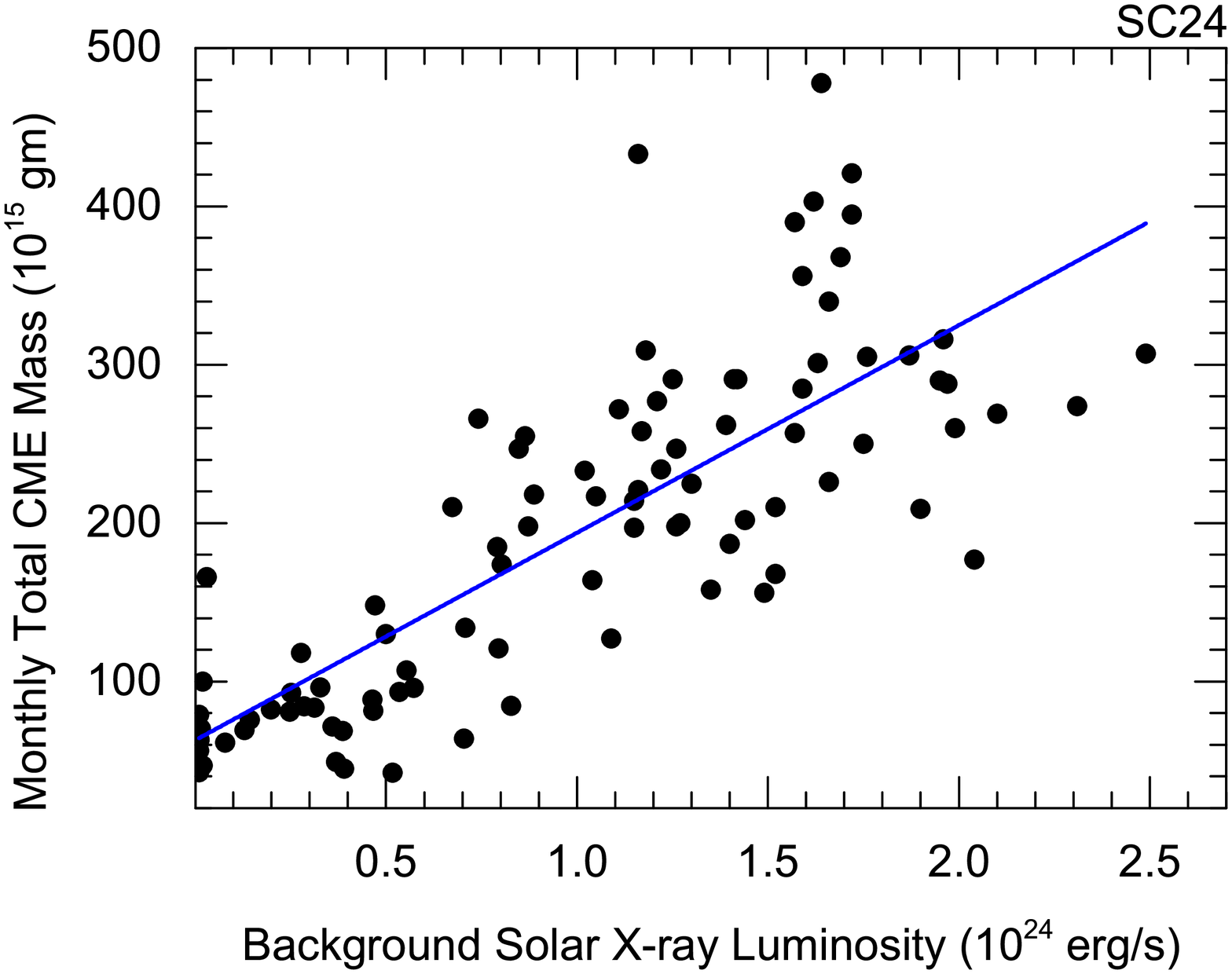}
		\caption{The variation of monthly mass loss (on the left Y-axis in black) due to all the CMEs from the Sun and  Solar X-ray background luminosity (on the right Y-axis in blue) with time (on X-axis) during 1996 to 2016 (i.e, solar cycle 23 to 24) is shown. The smoothed value of actual measurements are also overplotted (Top panel). The scatter plots between Solar X-ray background luminosity and monthly mass loss due to all the CMEs from the Sun with a fitted regression line (in blue) for solar cycle 23 (middle panel, labeled as SC23) and solar cycle 24 (bottom panel, labeled as SC24) is shown.}
	\label{CME_massloss_xr_SC2324}
\end{figure}

\begin{table}
  \caption{First panel from the top: The values of constants (c$_{1}$ and c$_{2}$), correlation coefficient ($r$) and coefficient of determination ($r^{2}$) obtained using Equation~\ref{eqcmlrss} (i.e., relation between monthly mass loss rate via three groups of the CMEs from latitudes within -30$^\circ$ to 30$^\circ$, -60$^\circ$ to 60$^\circ$, and -90$^\circ$ to 90$^\circ$ and monthly averaged sunspot number), for solar cycles 23 and 24. Second panel from the top: The values of c$_{1}$, c$_{2}$, $r$ and $r^{2}$ obtained using Equation~\ref{eqcmlrxr} (i.e., relation between monthly mass loss rate via all the CMEs from the Sun and solar X-ray background luminosity) for solar cycles 23 and 24. Third panel from the top: Similar to second panel, but using Equation~\ref{eqsmlrss} (i.e., relation between monthly mass loss rate via solar wind and monthly averaged sunspot number). Fourth panel from the top: Similar to second panel, but using Equation~\ref{eqsmlrxr} (i.e., relation between monthly mass loss rate via solar wind and solar X-ray background luminosity).}
  \label{coef_func}
  \begin{tabular}{m{2.2cm} m{1.7cm} m{0.8cm} m{0.8cm} m{0.7cm}}
  \hline \hline
	\multicolumn{5}{c}{Fitting from Equation~\ref{eqcmlrss} (CMEs and sunspot number)} \\ \hline 
	Cycle & c$_{1}$ & c$_{2}$ & r & r$^{2}$ (\%)\\ 	\hline
  SC 23 (-30$^\circ$ to 30$^\circ$) & 1.1  & 67   & 0.60  & 36.0   \\ 
  SC 23 (-60$^\circ$ to 60$^\circ$) & 2.4  & 85   & 0.72  & 51.8   \\
  SC 23 (-90$^\circ$ to 90$^\circ$) & 2.8  & 91   & 0.74  & 54.8    \\
	SC 24 (-30$^\circ$ to 30$^\circ$) & 1.2  & 82   & 0.57  & 32.5   \\ 
  SC 24 (-60$^\circ$ to 60$^\circ$) & 3.4  & 98   & 0.74  & 54.8    \\
  SC 24 (-90$^\circ$ to 90$^\circ$) & 4.1  & 118  & 0.76  & 57.8     \\  \hline
	
	\multicolumn{5}{c}{Fitting from Equation~\ref{eqcmlrxr} (CMEs and X-ray background luminosity)} \\ \hline
	 SC 23 & 2.5 $\times$ 10$^{-22}$  &  90   & 0.78  & 60.1   \\ 
	 SC 24 & 2.6 $\times$ 10$^{-22}$  &  126   & 0.80  & 64.0   \\  \hline
	
	\multicolumn{5}{c}{Fitting from Equation~\ref{eqsmlrss} (Solar wind and sunspot number)} \\ \hline
	 SC 23 & -0.2  &  696  & 0.1  & 1   \\ 
	 SC 24 & -0.1  &  616  & 0.03  & 0.09   \\  \hline
	
	\multicolumn{5}{c}{Fitting from Equation~\ref{eqsmlrxr} (Solar wind and X-ray background Luminosity)} \\ \hline
	 SC 23 & -1.2 $\times$ 10$^{-23}$  &  694   & 0.09  & 0.8   \\ 
	 SC 24 & 1.6 $\times$ 10$^{-23}$  &  594   & 0.09  &  0.8   \\  \hline
  \end{tabular}
  \end{table}

\subsection{Mass Loss Rate via CMEs Versus Solar X-ray Background Luminosity}
\label{cme_ml_sxr}
\subsubsection{Estimation of Solar X-ray Background Luminosity}

It is well-established that there is no one to one association between solar flares and CMEs, therefore, we associate the  CMEs mass loss rate with X-ray background luminosity ($L_{X}$) from the Sun. We investigate if $L_{X}$ is a better proxy than photospheric sunspot number for understanding CMEs mass loss rate with the solar cycle. Further, measuring X-ray luminosity is important as it has a high dynamic range between solar maximum and minimum \citep{Mewe1972,Aschwanden1994} and also is responsible for disturbances in the planetary atmosphere \citep{Tsurutani2009}.

The soft X-ray background flux from the Sun can be calculated by subtracting out the contribution of solar flares. For this purpose, we used a robust method as described in \citet{Cohen2011}. In this method, the measured GOES X-ray flux higher than 10$^{-6}$ W m$^{-2}$ is removed from the data. This is done as the typical non-flaring X-ray fluxes measured in GOES ranges between 5 $\times$ 10$^{-9}$ to 10$^{-7}$  W m$^{-2}$, thus the typical measured GOES flux associated with a solar flare is much higher than 10$^{-6}$ W m$^{-2}$ \citep{Bornmann1990,Veronig2004,Golub2009}. After excluding the luminosity contributed by flare data points, we calculated the monthly average of the non-flare GOES X-ray flux. We then converted the measured X-ray background flux into the X-ray luminosity from the Sun. To keep our analysis in the astronomical context where CGS units are used, the X-ray luminosity is converted into the unit of erg s$^{-1}$. This X-ray background luminosity (i.e., quiescent X-ray luminosity) would be governed by free-free bremsstrahlung and line emission of heated plasma confined in active regions by closed magnetic loops \citep{Aschwanden1994}.

\subsubsection{CME Mass Loss Rate and Solar X-Ray Background Luminosity During Solar cycles 23 and 24} 
\label{cme_ml_sxr_main}

The variation in mass loss rate via all the CMEs and solar X-ray background luminosity during solar cycle 23 and 24 is shown in Figure~\ref{CME_massloss_xr_SC2324}. We note that the soft X-ray background luminosity varies by a factor of 
10 between solar maximum and minimum. From the top panel of the figure, we note that the variation in mass loss rate follows the variations in solar X-ray background luminosity in general. However, it can be seen that during the rising and maximum phase of the solar cycle 23 from the year 1997 to 2003, the amplitude of mass is relatively lower than X-ray luminosity. The similar pattern is also noted during the maximum of solar cycle 24. The X-ray luminosity and the mass loss per month at the maximum of cycle 24 is decreased by around 20\% and 15\%, respectively, than that during the maximum of the previous cycle 23. This implies almost an equal reduction of mass loss rate due to CMEs and X-ray background luminosity in contrast to sunspot number which is decreased by around 40\% from cycle 23 to 24. We note cyclical activity on a similar time scale in both the X-ray background and sunspot number.

We fitted the estimated mass loss rate via CMEs ($dM_{CME}/dt$) anywhere from the latitudes within -90$^\circ$ to 90$^\circ$ of the Sun and X-ray background luminosity ($L_{X}$) with a mathematical function given as,

\begin{equation}
 \frac{dM_{CME}}{dt}=5\times 10^{14} (c_{1} L_{X} + c_{2})    \text{~~~~gm month$^{-1}$}
	\label{eqcmlrxr}
\end{equation}

where c$_{1}$ and c$_{2}$ are the constants. The obtained values of constants, correlation coefficient and coefficient of determination for cycles 23 and 24 are shown in the second panel from the top of Table~\ref{coef_func}. From the table, we note that Equation~\ref{eqcmlrxr} could explain around 60\% and 64\% of the variability in mass loss rate for solar cycle 23 and 24, respectively. The middle and bottom panels of Figure~\ref{CME_massloss_xr_SC2324} show scatter plots of the monthly CME mass loss rate as a function of the X-ray background luminosity for solar cycles 23 and 24, respectively. The variability in mass loss rate explained by X-ray background luminosity is around 5\% higher for both the cycles than that using the monthly averaged sunspot number. This implies that measured X-ray luminosity is a better proxy for the mass loss activity due to CMEs for the whole Sun. Interestingly, the sunspot cycle 23 and X-ray luminosity peaks at the same time around mid-2001. However, the sunspot cycle 24 peaks at September 2013 while the X-ray luminosity peak is delayed by half-year. We also note that X-ray background flux and mass loss rate relative to the number of sunspots is higher for sunspot cycle 24 than cycle 23. The CME mass loss rate which is largely contributed by strong events unlikely to be missed in CDAW CME catalog, it is obvious that there is an enhanced CME activity since the middle of solar cycle 23 and continues during weaker sunspot cycle 24. Our finding is in contrast to \citet{Lamy2014} which shows that CME activity is well correlated with the sunspot number during both the cycles 23 and 24.

\section{Variability in Solar Wind Mass Flux and Resulted Mass Loss Rate During Solar Cycles 23 and 24}
\label{sw_ml_ssxr}

We used solar wind measurements from in situ instruments on board WIND spacecraft located in the ecliptic at around 1 AU distance from the Sun. The solar wind data was obtained from CDAWeb (\url{https://cdaweb.sci.gsfc.nasa.gov/cgi-bin/eval2.cgi}) website. We make it clear that by referring the term in situ observations of "solar wind" we mean the in situ observations of both the quasi-steady solar wind and episodic mass ejections from the Sun. Thus, the term mass loss rate via solar wind, as used in our study, refers to the rate of the total mass expelled from the Sun into the heliosphere. The contribution of mass via CMEs into the solar wind is estimated in Section~\ref{cme_sw_rat}.

\begin{figure}
	\centering
		\includegraphics[scale=0.30]{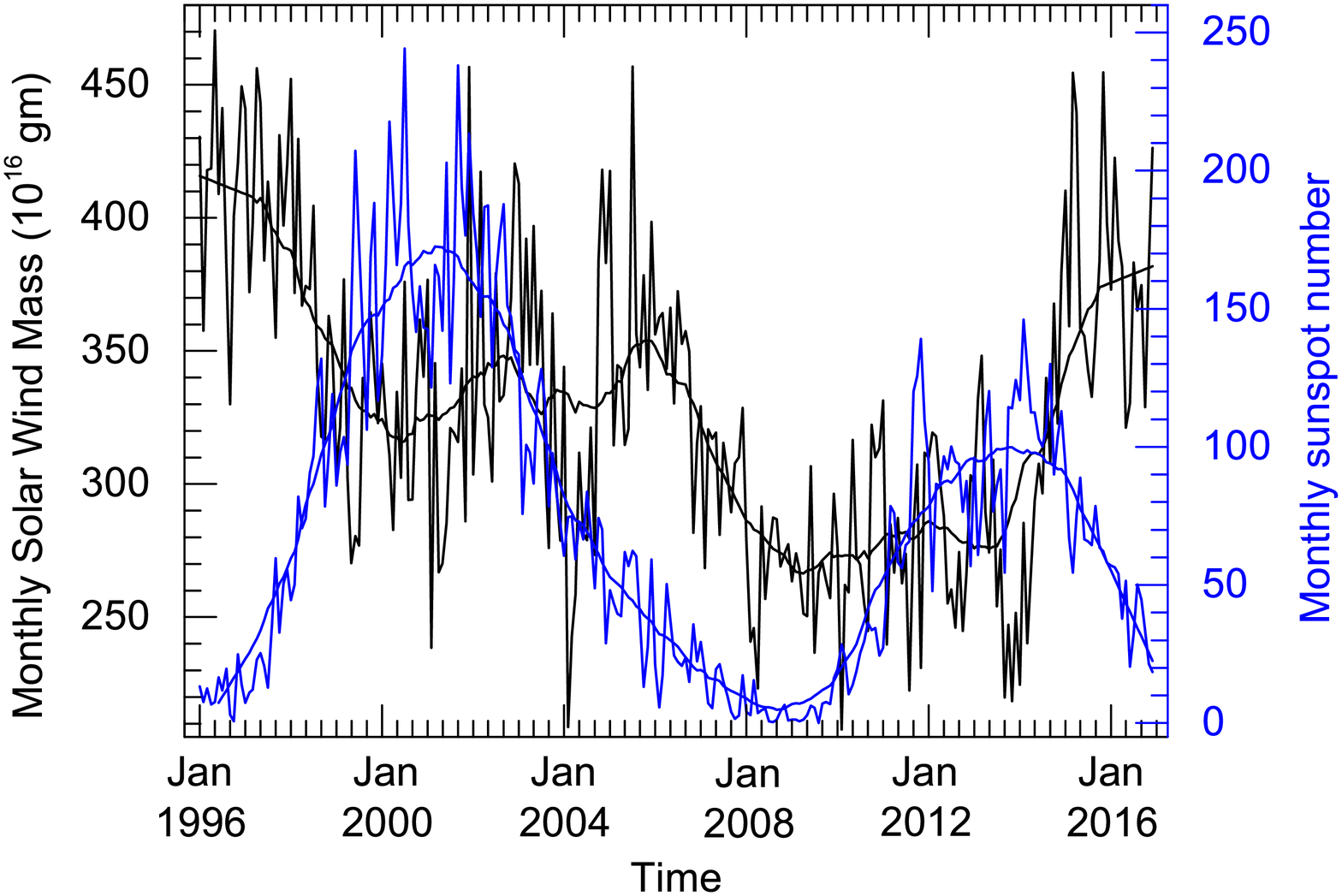} 
		\vspace{2mm}
		\includegraphics[scale=0.30]{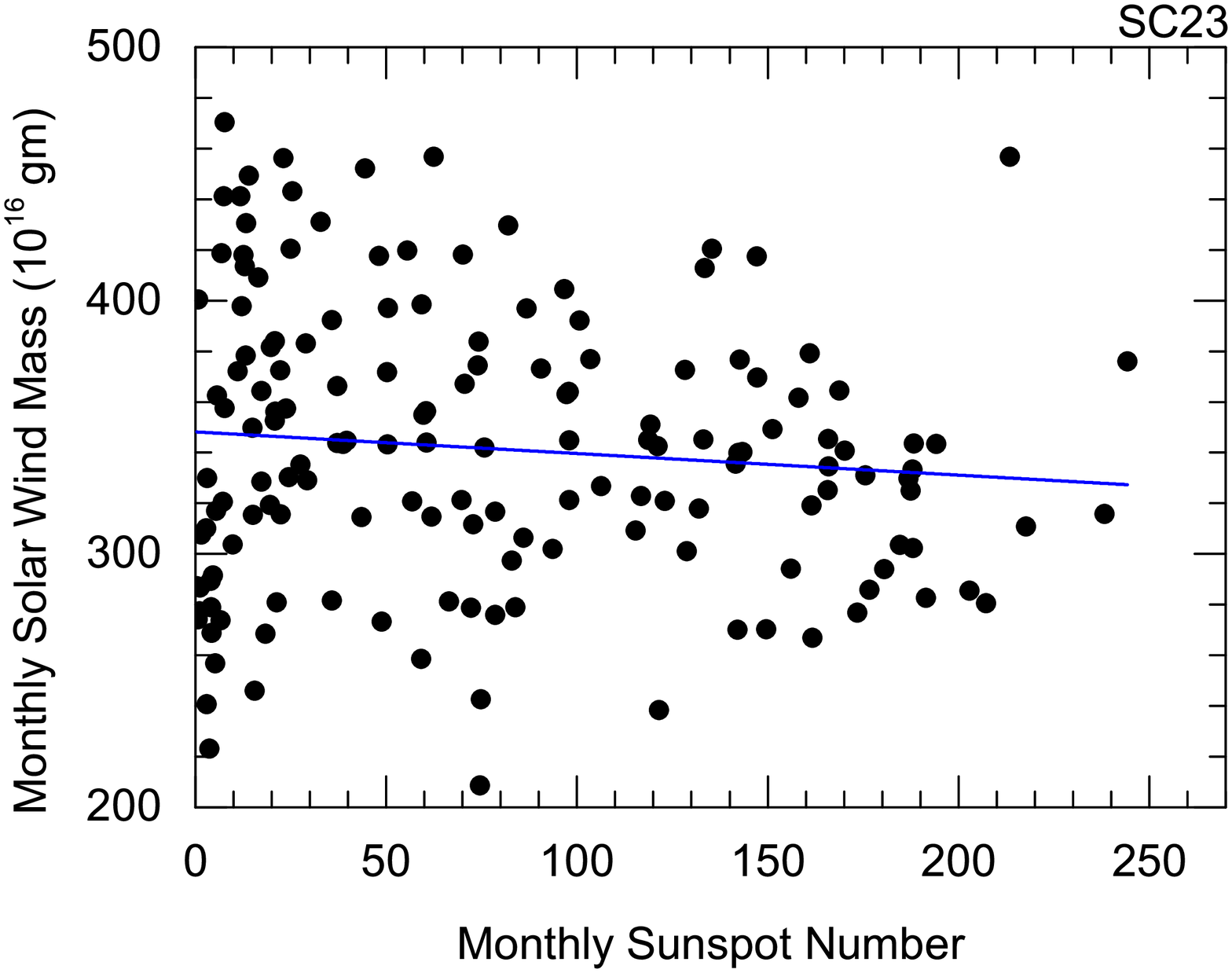}
		\vspace{2mm}
		\includegraphics[scale=0.30]{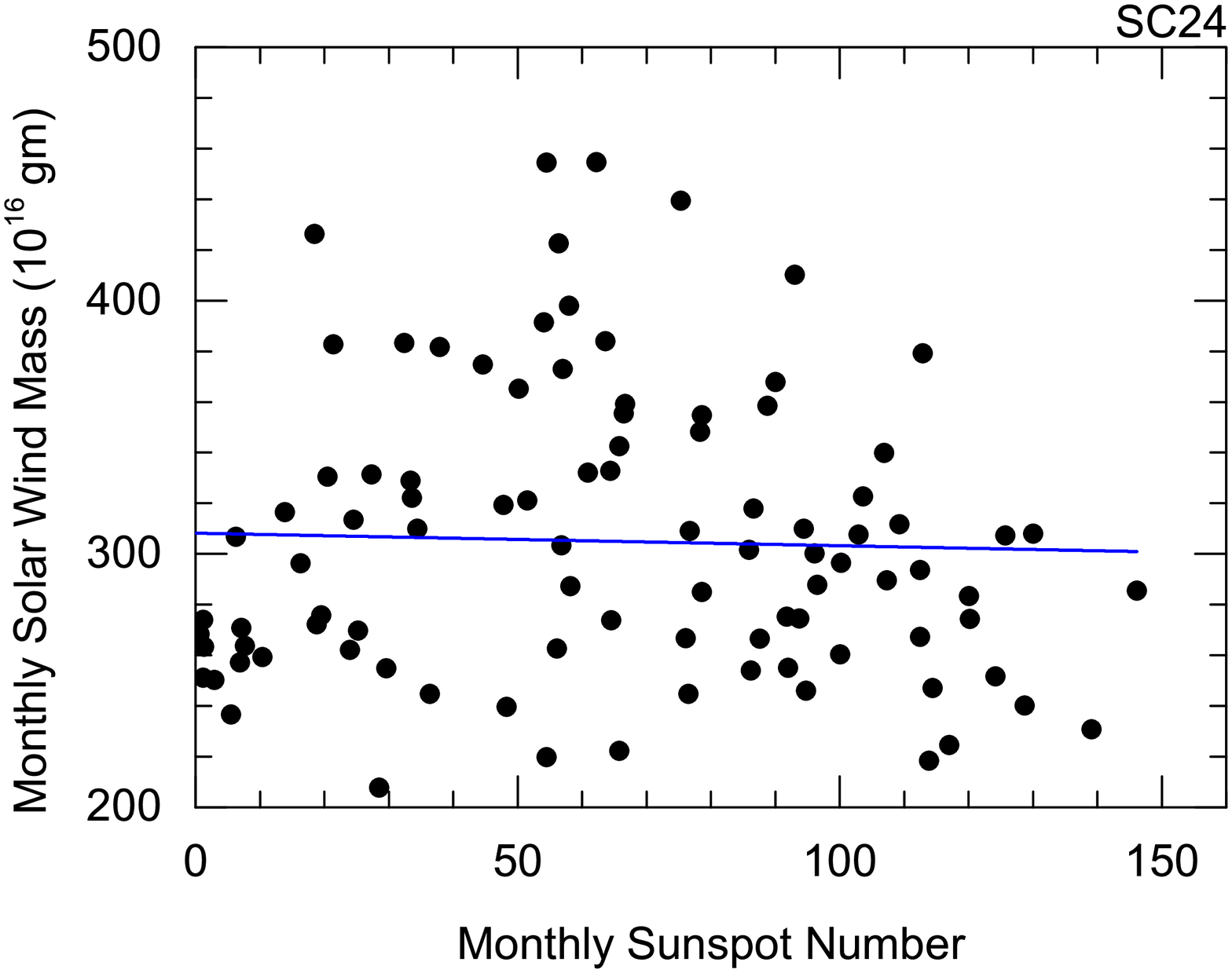}
		\caption{The variation of monthly mass loss (on the left Y-axis in black) due to solar wind and monthly Sunspot number (on the right Y-axis in blue) with time (on X-axis) during 1996 to 2016 (i.e, solar cycle 23 to 24) is shown. The smoothed value of actual measurements are also overplotted (Top panel). The scatter plots between monthly sunspot number and monthly mass loss due to solar wind with a fitted regression line (in blue) for solar cycle 23 (middle panel, labeled as SC23) and solar cycle 24 (bottom panel, labeled as SC24) is shown.}
	\label{sw_massloss_ss_SC2324}
\end{figure}

\subsection{Mass Loss Rate via Solar Wind Versus Sunspot Numbers}
\label{sw_ml_ss}

Using the hourly averaged in situ observations of solar wind near 1 AU, we estimated the monthly averaged proton density 
($n_{p}$), speed ($v_{p}$) and resulting proton mass flux ($n_{p}$ $\times$ $v_{p}$) at 1 AU. Considering the measured solar wind mass flux as the global solar wind mass flux at 1 AU in all directions from the solar surface, we determined the monthly solar wind mass loss rate. The variation in the estimated monthly mass loss via solar wind as a function of monthly averaged sunspot number for sunspot cycles 23 and 24 is shown in Figure~\ref{sw_massloss_ss_SC2324}. From the top panel of the figure, we note that the monthly mass loss decreases during the rising and maximum phase of solar cycle 23. However, in the declining phase of cycle 23, the mass loss rate slightly increases in the beginning and then decreases tracking the sunspot number to some extent. For solar cycle 24, the solar wind mass loss rate remains almost constant during the rising phase and decreases slightly during the maximum and increases during the declining phase.

We fitted the estimates of mass loss rate and sunspot number using Equation~\ref{eqsmlrss}, 

\begin{equation}
 \frac{dM_{SW}}{dt}=5\times 10^{15} (c_{1} S + c_{2})    \text{~~~~gm month$^{-1}$}
	\label{eqsmlrss}
\end{equation}   

where $S$ is the monthly averaged sunspot number, c$_{1}$ and c$_{2}$ are the constants. The middle and bottom panels of Figure~\ref{sw_massloss_ss_SC2324} shows a scatter plot of the monthly mass loss rate as a function of the sunspot number for solar cycles 23 and 24, respectively. The value of the constants, correlation coefficient and coefficient of determination are given in the third panel from the top of Table~\ref{coef_func}. From the table, we note that the variability in the mass loss via solar wind could not be properly explained in terms of a linear function of sunspot number as given in Equation~\ref{eqsmlrss}. The solar wind mass loss rate is scattered around an average value 3.5 $\times$ 10$^{18}$ gm month$^{-1}$ for cycle 23 and the value decreases by only around ~10\% for cycle 24. The sunspot number and CME mass loss rate at the maximum of cycle 24 decreased by approximately 40\% and 15\%, respectively, than that during the maximum of cycle 23. Also, the intra-cycle variations in the mass loss rate via solar wind seems to be different for solar cycles 23 and 24.

\begin{figure}
	\centering
		\includegraphics[scale=0.30]{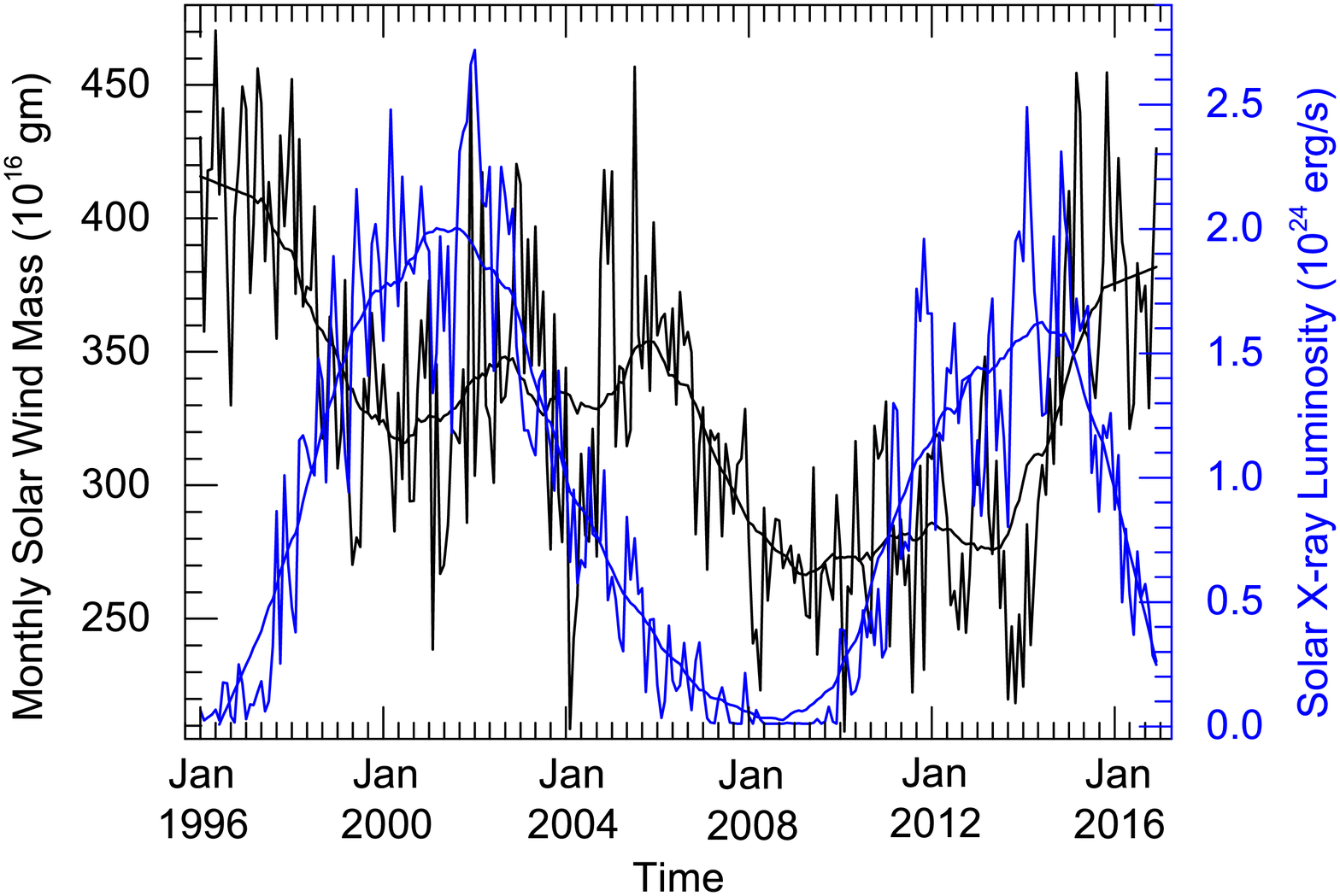} 
		\vspace{2mm}
		\includegraphics[scale=0.30]{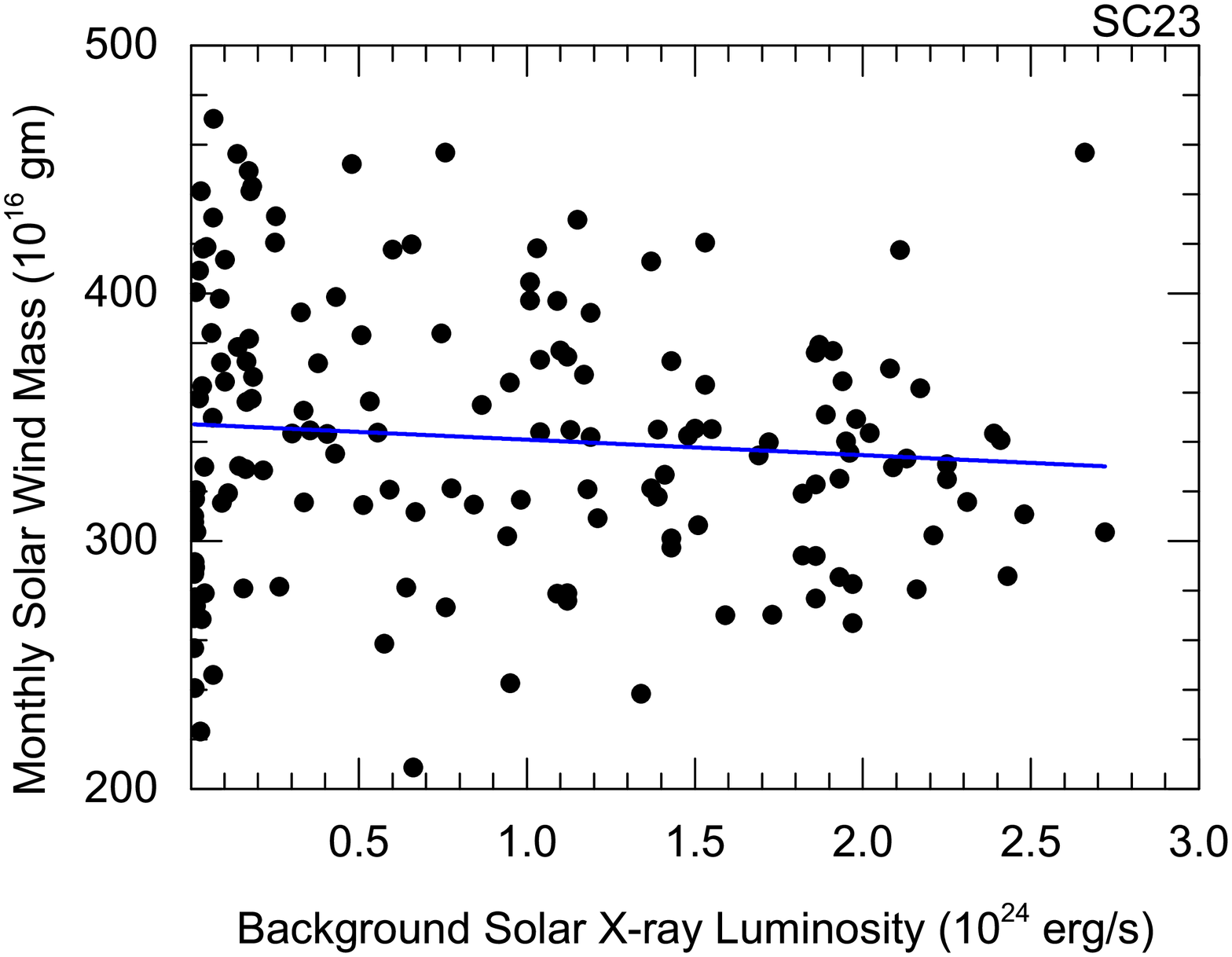}
		\vspace{2mm}
		\includegraphics[scale=0.30]{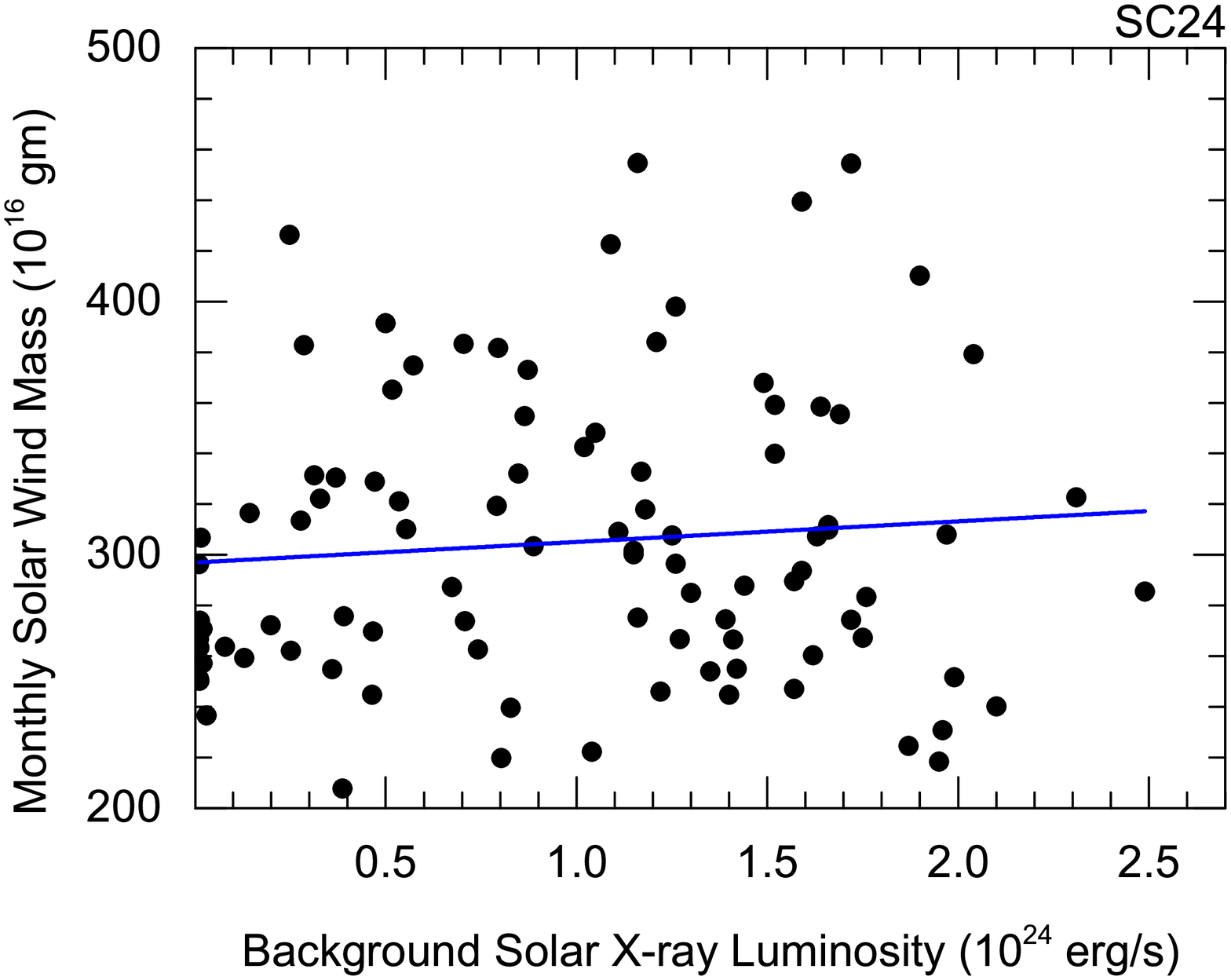}
		\caption{Similar to Figure~\ref{sw_massloss_ss_SC2324}, but background solar X-ray luminosity is shown instead of monthly sunspot number.}
	\label{sw_massloss_xr_SC2324}
\end{figure}

\subsection{Mass Loss Rate via Solar Wind Versus Background Solar X-rays Flux}
\label{sw_ml_xr}

The variation in the monthly mass loss via solar wind with solar X-ray background luminosity is shown in 
Figure~\ref{sw_massloss_xr_SC2324}. The Solar X-ray background luminosity closely tracks the sunspot cycles. From the top panel of the figure, we note that the  mass loss rate decreases during the rising and maximum phase of the solar cycle 23.  During the declining phase of cycle 23, the monthly mass loss slightly increases in the beginning and then tracks the X-ray luminosity. For solar cycle 24, the mass loss rate is almost constant in the beginning and then decreases during the rising phase of cycle 24. The mass loss rate increases during the maximum and declining phase of cycle 24. We note that that mass loss rate via solar wind neither follows the solar cycles trend nor shows a significant variation from the maximum to minimum of the cycles while X-ray background luminosity varies by an order of magnitude.

To look into general nature of mass loss rate via solar wind and X-ray luminosity, we fit them using Equation~\ref{eqsmlrxr}, 

\begin{equation}
 \frac{dM_{SW}}{dt}=5\times 10^{15} (c_{1} L_{X} + c_{2})    \text{~~~~gm month$^{-1}$}
	\label{eqsmlrxr}
\end{equation}

where $L_{X}$ is the monthly averaged solar X-ray background luminosity, c$_{1}$ and c$_{2}$ are the constants. The best fit of the data for solar cycle 23 and 24 is shown in the middle and bottom panels of Figure~\ref{sw_massloss_xr_SC2324}, respectively. The constants, correlation coefficient and coefficient of determination derived from the fitting are shown in the bottom panel of Table~\ref{coef_func}. We note that variance in the mass loss rate is not accounted for by independent covariate (i.e., $L_{X}$) used in the mathematical function. The modulations in the solar X-ray background luminosity have not an obvious and significant effect on the mass loss rate via the solar wind. The possible explanation of relatively larger independency of solar wind mass loss rate on the proxy of solar activity (i.e., sunspot numbers and X-ray luminosity) is discussed in Section~\ref{resdis}.

\begin{figure}
	\centering
		\includegraphics[scale=0.47]{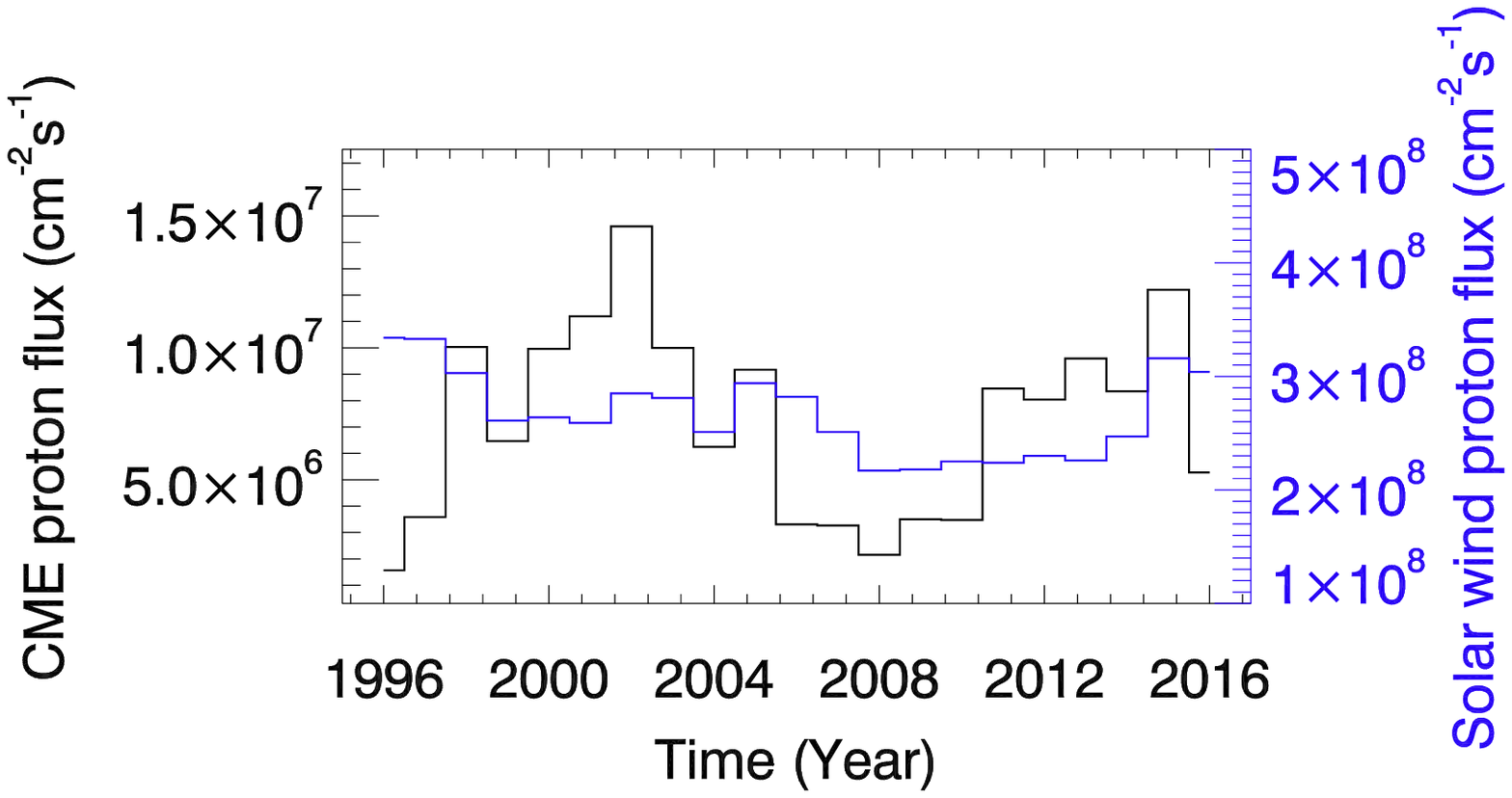} \\ 
		\vspace{3mm}
		\includegraphics[scale=0.50]{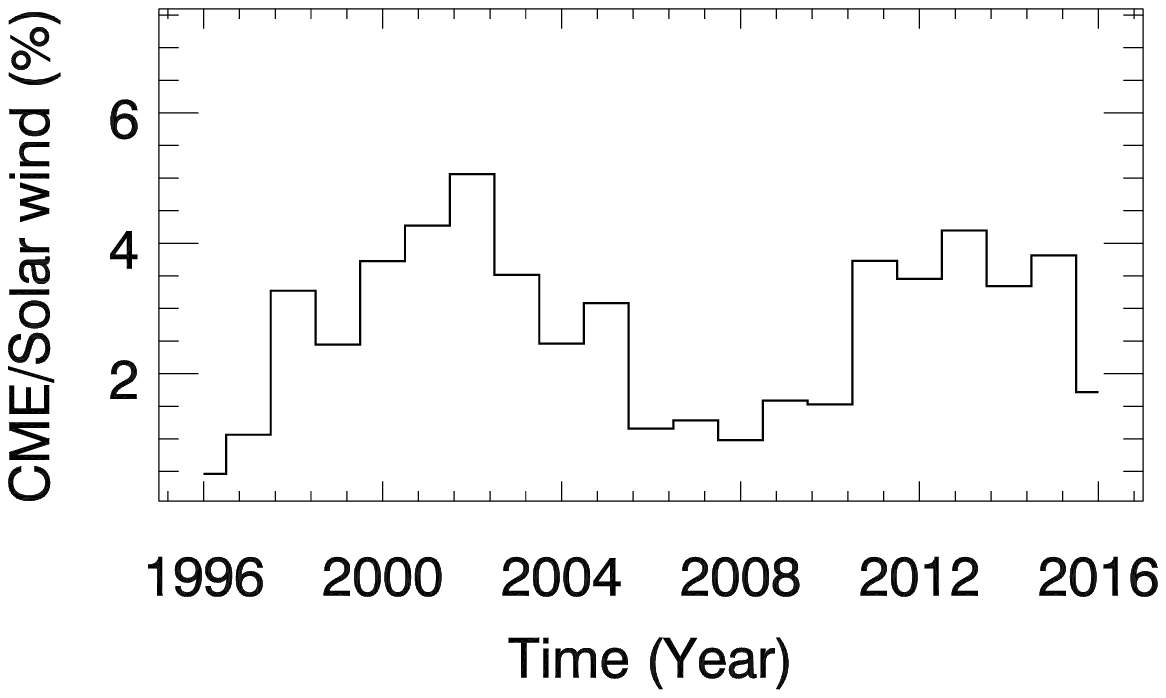}
		\caption{Top panel: The variation of CMEs proton flux (on the left Y-axis) and solar wind proton flux (on the right Y-axis in blue) at 1 AU in the near-ecliptic is shown with time (on X-axis). Bottom panel: The ratio of CME to solar wind mass flux is shown.}
	\label{cme_sw_r}
\end{figure}

\subsection{Relative Contribution of CMEs to the Ambient Solar Wind}
\label{cme_sw_rat}
The mass loss rate via solar wind, as calculated using Equations~\ref{eqsmlrss} and \ref{eqsmlrxr}, constitutes the mass loss via quasi-steady solar wind as well as episodic ejections from the Sun. The quasi-steady solar wind comes from all over the solar surface while most of the CMEs are from low-mid latitudes active regions. We attempt the estimate the contribution of CMEs to the solar wind mass flux measured at 1 AU near the Earth. We also examine how the relative contribution of CMEs varies over the different phases of solar cycles 23 and 24.

\subsubsection{Relative Contribution of CMEs to the Solar Wind in the Ecliptic}
\label{cmeswr_ecl}
CMEs have certain angular width and therefore they shed mass over a wider heliolatitudinal range. We assume that the mass of a CME to be distributed uniformly within its span. Taking into account the latitudes of CMEs, their angular sizes and their cumulative contribution along the ecliptic, we estimated the annually averaged CME mass ejected per day into a 
1$^\circ$ wide equatorial latitude bin. The annual average was taken to make a reasonable distribution of mass over all the latitudinal bins and will also facilitate us to compare our results with earlier studies \citep{Howard1985,Lamy2017}. We can assume that the annually averaged CME mass ejected per day into a 1$^\circ$ wide equatorial latitude bin is distributed uniformly in 90$^{\circ}$ wide longitudinal sectors centered on both the east and west limbs. We also consider that mass ejected from the Sun along the ecliptic remains almost in the same plane during its heliospheric propagation. To determine the equatorial mass flux at 1 AU, we divide the mass ejected into equatorial bin by 2$\pi$ $r$ $h$, where $r$=1 AU and $h$ = 2.6 $\times$ 10$^{11}$ cm is an arc length of 1$^\circ$ wide latitudinal sector at a radial distance of 1 AU. Further, assuming a composition of CME in which helium constitutes around 10\% and hydrogen is around 90\%, i.e., around 69.5\% of the total mass is due to protons, we determined the proton flux at 1 AU due to the CMEs. We also derived the annual average of solar wind proton mass flux at 1 AU based on the in situ observations. The solar cycle variation of solar wind proton flux, CME proton flux and the contribution of CMEs to the solar wind at 1 AU in near-ecliptic are shown in Figure~\ref{cme_sw_r}. We note that that solar wind mass flux is almost constant over time, and the contribution of CMEs to the solar wind is negligibly small during the solar minimum but increased to $\approx$5\% at the maximum of solar cycles 23 and 24.

\subsubsection{Relative Contribution of CMEs to Solar Wind Mass Fluxes at Different Heliographic Latitudes}

\begin{figure}
	\centering
		\includegraphics[scale=0.62]{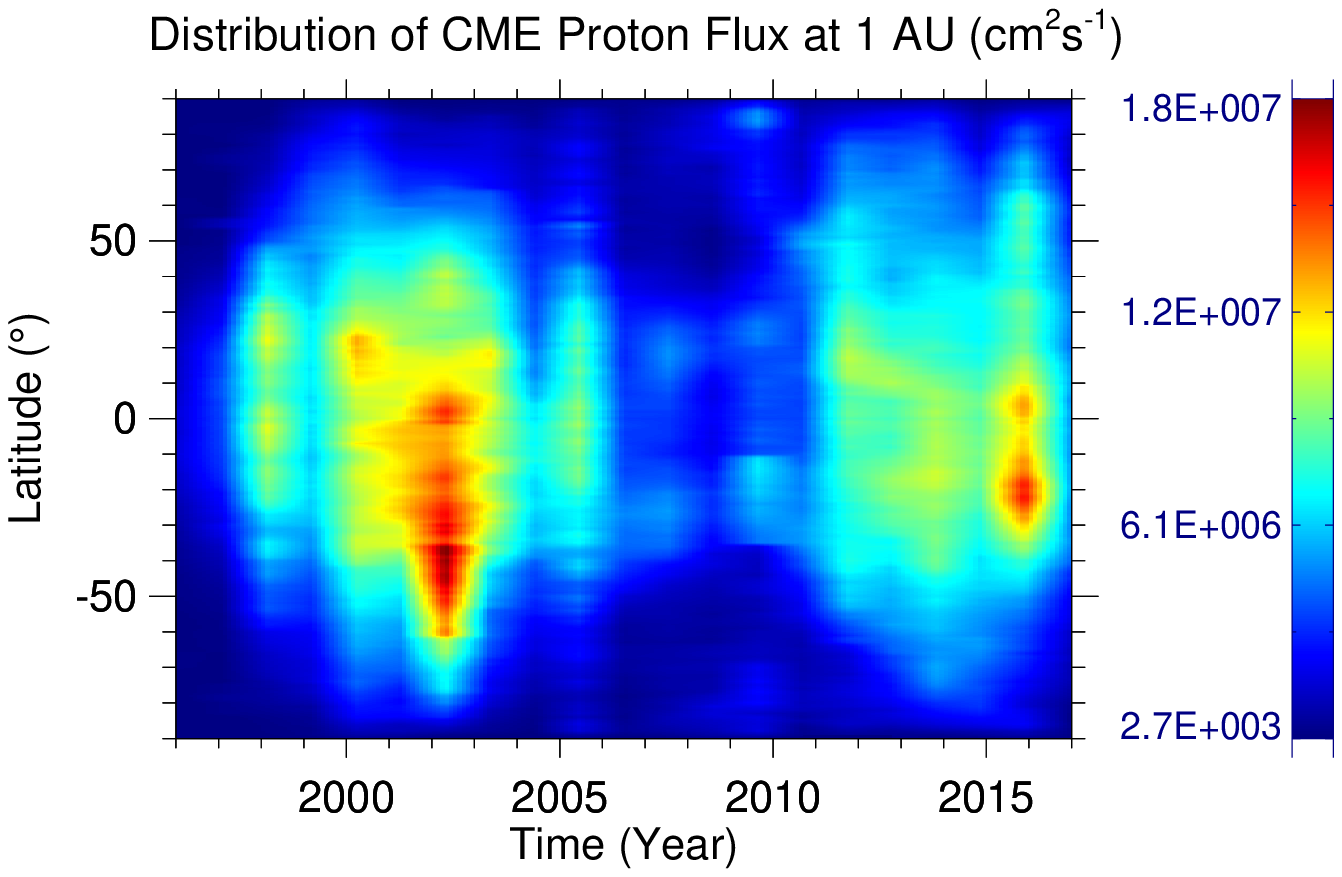} \\ 
		\vspace{2mm}
		\includegraphics[scale=0.45]{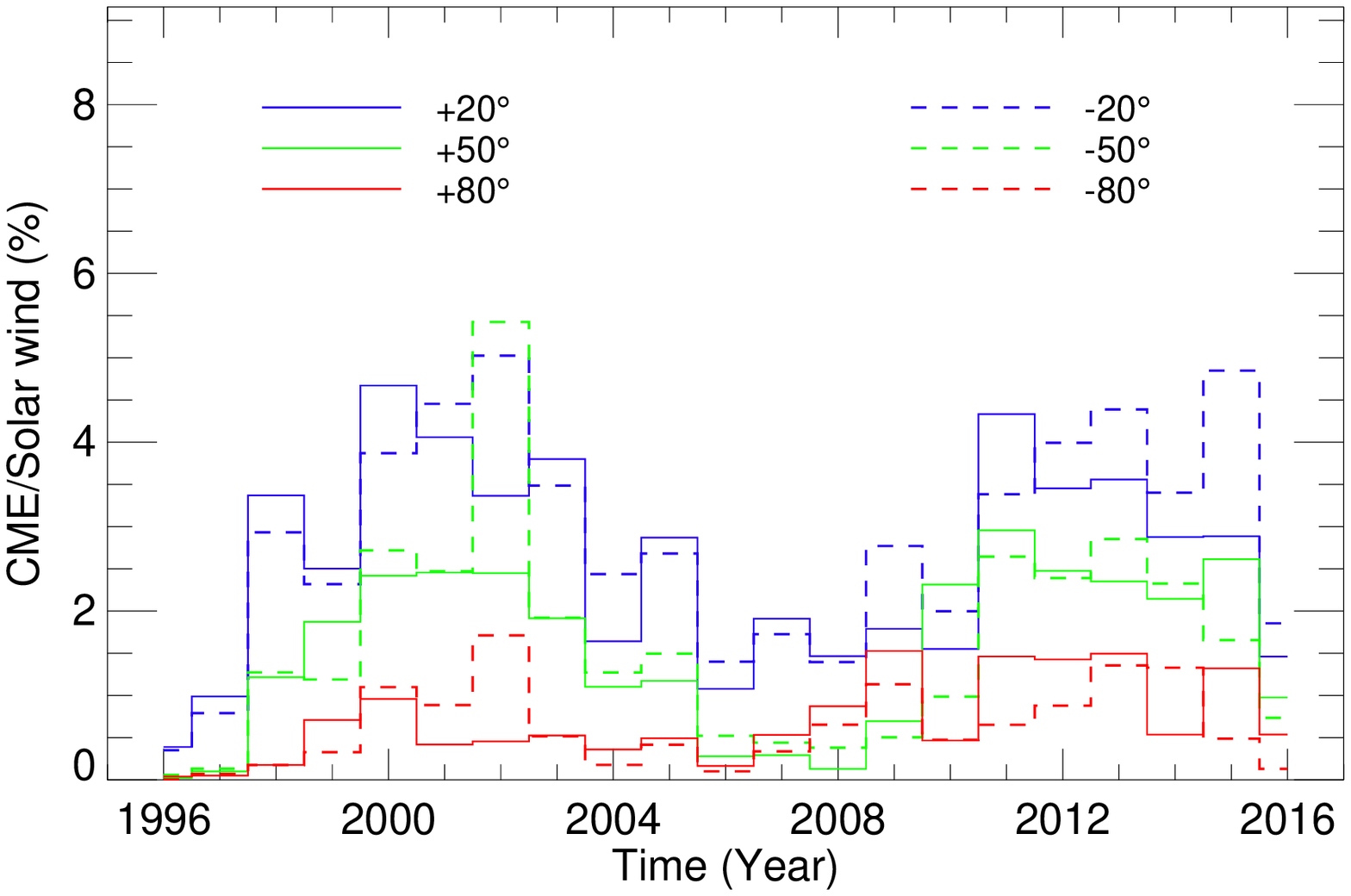}
		\caption{Top panel: the distribution of CME proton flux at 1 AU over all the heliolatitudes for solar cycle 
		23 and 24 is shown. Bottom panel: the ratio of CME to solar wind mass flux along the different heliolatitudes 
		is shown.}
	\label{cmeswr_lat}
\end{figure}

Following the approach described in Section~\ref{cmeswr_ecl}, we estimated the heliolatitudinal distribution of annually averaged CME mass per day per degree. Using the near Sun estimates of mass flux, we obtained the latitudinal variation of CME proton flux and the relative contribution of CMEs to total solar wind at 1 AU for solar cycles 23 and 24 as shown in Figure~\ref{cmeswr_lat}. We note that during the solar maximum, larger amount of CME mass tends to reach at upper latitudes making a dense mass flux distribution at low, mid and high latitudes. However, during the solar minimum, a lesser CMEs mass is distributed along the equatorial and polar latitudes which make a denser mass distribution along the mid-latitudes. Further, mass flux was observed to increase extending up to higher latitudes for a longer duration from 2000 to 2004 during the maximum of solar cycle 23 than that during the maximum of solar cycle 24. This implies that mass loss via CMEs from solar cycle 24 is smaller than that of the previous cycle. We also note that mass flux distribution reaches to higher latitudinal bins for solar cycle 24 than the previous cycle. Both the cycles have larger mass flux in the southern hemisphere than that in northern hemisphere suggesting an asymmetry in the mass flux in the heliosphere.

Assuming that mass flux via quasi-episodic solar wind remains almost the same over the complete solar surface, we calculated the relative contribution of CMEs into the solar wind at different heliolatitudes as shown in the bottom panel of Figure~\ref{cmeswr_lat}. It is noted that irrespective the heliolatitudes considered, the relative contribution of CMEs to the solar wind mass flux is small near the solar minimum and increases as the cycle rises. The fractional contribution of CME to solar wind peaks around 2 years before the  maximum of sunspot cycles 23 and 24 for all the northern heliolatitudes. However, for all the southern heliolatitudes, the contribution of CMEs peaks around the maximum of sunspot cycle 23 and around 2 years after the maximum of sunspot cycle 24. The amplitude of the peak along the southern latitudes is slightly larger than that along the northern latitudes. In general, the fractional contribution of CMEs becomes smaller along the higher heliographic latitudes. The exception exists within the southern latitudinal range of -20$^\circ$ to -50$^\circ$ during the maximum of cycle 23 and after 2 years of the maximum of cycle 24. During the maximum of the cycles, the fractional contribution of CMEs to the solar wind is around $\approx$ 5\% along the near ecliptic and becomes $\approx$ 1\% along the 80$^\circ$ of heliolatitude.

\section{Results and Discussion}
\label{resdis}

\subsection{CME Mass Loss Rate and Proxies of Solar Cycle}

We studied the solar cycle dependence in the occurrence rate of CMEs, ICMEs, mass loss rate via CMEs originating from different solar latitudinal bins, and mass loss via solar wind with different proxies of solar cycle over the period of 1996 to 2016 covering the solar cycle 23 and 24. We established a relationship between monthly CME mass loss rate and monthly sunspot number, and the correlation coefficient between them is found to be 0.74 and 0.76 for the cycle 23 and 24, respectively. A similar solar cycle dependence of CME mass loss rate was observed in the study of \citet{Cranmer2017} where the correlation coefficient between CME mass loss rate and fitting function was found to be 0.81 based on the CDAW CMEs during 1996 to 2013. A good correlation between sunspot number and mass loss rate suggests that they are probably the consequences of a common magnetic activity cycle. We find an average mass loss rate $dM_{CME}/{dt}$ = 3 $\times$ 10$^{-18}$ (2.8 $S$ + 91) M$_\odot$ yr$^{-1}$ for cycle 23 and $dM_{CME}/{dt}$ = 3 $\times$ 10$^{-18}$ (4.1 $S$ + 118) M$_\odot$ yr$^{-1}$ for cycle 24. We also note that the sunspot number explains the variability in the mass loss much better when the CMEs from wider latitudinal range is considered (Table~\ref{coef_func}). Thus, the sunspot number can be used as a predictor of the CME mass loss rate from all the latitudes. We deduce that the latitudinal distribution of CME mass loss rate, or indeed their latitudinal distribution is different from those of sunspot groups (Figure~\ref{CME_massloss_ss_SC2324}). It means that irrespective of the phases of a solar cycle, a fraction of mass loss is also from higher latitudes probably originating from quiescent filament region. It is noted that around 10\% and 20\% of the total CME mass loss during the rising phase of the solar cycle 23 and 24, respectively, is contributed from CMEs having their apparent source latitudes higher than $\pm$60$^\circ$. The fraction of total CME mass loss via CMEs from higher latitudes is reduced by a factor of 2 during the declining phase of solar cycles. This further confirms that active sunspot regions represent only a subset of the source region of the CMEs \citep{Hundhausen1993,Gopalswamy2003,Robbrecht2009}.

Further, the CME mass loss rate in connection with soft X-ray background luminosity is explored, and the correlation coefficient between them is found to be 0.78 and 0.80 for cycle 23 and 24, respectively. The background X-ray luminosity seems to be a better predictor of CME mass loss rate. This may be because the source of soft X-ray is in corona from where CMEs are launched while the sunspot number is a good representative of activity at the photosphere. Thus, the number of sunspots may suggest for the magnetic field generated by solar dynamo while the X-ray luminosity may suggest for transportation of magnetic flux to the corona. The better correlation between CME activity and background X-ray luminosity may also be because CMEs originate not only from low latitude active regions containing sunspots but also from high latitude non-sunspot  regions containing filaments which emit in soft X-ray. This is also suggested from our analysis of CME mass loss rate over different latitudinal bins as shown in Figure~\ref{CME_massloss_ss_SC2324}. We noted that X-ray background luminosity and sunspot number follows the variation in mass loss rate via all the CMEs quite well, and thus a linear relationship between them can be found. We estimate an average mass loss rate $dM_{CME}/{dt}$ = 3 $\times$ 10$^{-18}$ (2.5 $\times$ 10$^{-22}$ $L_{X}$ + 90) M$_\odot$ yr$^{-1}$ for cycle 23 and $dM_{CME}/{dt}$ = 3 $\times$ 10$^{-18}$ (2.6 $\times$ 10$^{-22}$ $L_{X}$ + 126) M$_\odot$ yr$^{-1}$ for cycle 24. Our study suggests that X-ray background luminosity may be used as a primary scaling variable for determining stellar CME properties. Further, it would be interesting to estimate the X-ray luminosity and CME mass loss rate during historical time periods using the historical sunspot observations.

We find that solar activity measured in terms of sunspot number is weaker for solar cycle 24 than the previous cycle, and it did not quite translate into occurrence rate of CME in CDAW catalog that uses both LASCO C2 and C3 images. During the maximum of cycle 24, the sunspot number is reduced by 40\%, the rate of CME is increased by 60\% while the rate of ICMEs remains almost the same than that during the maximum of cycle 23. Thus, a larger number of CMEs and ICMEs per sunspot number is noted for cycle 24 than cycle 23. The occurrence rate of CME in the CDAW catalog is increased by a factor of 10 and 4 from minimum to maximum of the solar cycle 23 and 24, respectively. We noted that the CME occurrence rate from the CDAW C2+C3 catalog has increased since the year 2003 to 2016, however, a similar increase is not obvious for ICME rate. In a few earlier studies, the increased occurrence rate of CME derived from SEEDS catalog that uses LASCO C2 images only, has been attributed to the doubling of LASCO image cadence since increased telemetry became available from late 2010 onward \citep{Wang2014a, Hess2017}. They showed that cadence corrected CME rate from SEEDS C2-only catalog was well correlated with the sunspot number. However, we think that the cadence correction made by \citet{Wang2014a} was excessive. This is because a vast majority of detected CMEs are found in several successive LASCO images and   there is a lack of evidence for increased rate of fast and faint CMEs only detectable in the high-cadence sequence of images. Therefore, halving the image rate should not eliminate half of the CMEs detection. Further, we note the increased rate of CMEs in the CDAW catalog around 2003 which almost 7 years before the change in LASCO image cadence. Therefore, contrary to \citet{Wang2014a}, we suggest that a change in LASCO image cadence in 2010 may have only a little effect on the increase in CME activity during solar cycle 24.

To understand the CME activity over solar cycles 23 and 24, we estimated CME mass loss rate which is largely governed by wider and massive CMEs which are unlikely to be missed in automated and manual CME catalogs. Our finding suggests that a major fraction of CME mass loss can be accounted in terms of CMEs originating from lower-mid latitudes. However, a small fraction such as around 10\% and 20\% of the total CME mass loss is from latitudes beyond -60$^\circ$ to 60$^\circ$ during the maximum of the solar cycle 23 and 24, respectively. From the obtained mass loss rate as shown in Figure~\ref{CME_massloss_ss_SC2324}, it is obvious that latitudinal distribution of CMEs in both the cycles is different, and CME mass loss from higher latitudes is relatively larger in cycle 24 than that in the previous cycle \citep{Gopalswamy2015}. Importantly, the CME mass loss rate during the maximum of solar cycle 24 is decreased by around 15\% than that during the previous maximum. The CME mass loss rate in comparison to sunspot number is increased since the year 2003 which is similar to, but a much stronger, increase in occurrence rate of CDAW CME. It implies that the rate of less massive CMEs from higher latitudes, tend to begin outside of active regions, has resulted in the higher CME occurrence rate during a weaker sunspot cycle 24 since the year 2003, and it also has an appreciable contribution to the mass loss process. Our study suggests that a relatively lower activity in terms of the sunspot number may not necessarily reduce the CME activity to the same extent. Our study shows that there is a true physical increase in CME mass loss rate relative to the sunspot number in cycle 24.

Our finding of relatively increased mass loss rate from higher latitudes since 2003 is in agreement to earlier studies of \citet{Luhmann2011,Petrie2015} where the weakening of polar photospheric magnetic field which began around 2003 seems to be responsible for increased occurrence rate of CMEs since then. Our result of increased CME activity measured in terms of mass loss rate is in agreement to earlier studies which used C2+C3 catalogs (e.g., CDAW and CACTus) and they showed a real increase in CME occurrence rate around 2003 \citep{Gopalswamy2015,Petrie2015}. We differ from earlier studies which were based on the C2-only catalogs (e.g., SEEDS and ARTEMIS) and they either have shown an increase in CME occurrence rate from 2010 as an artifact \citep{Wang2014a} or no significant increase in CME rate \citep{Lamy2014}. Further investigation is required if the restriction of SEEDS detection to the C2 field of view overlook increased CME rate around the  middle of cycle 23 but notices it at the beginning of cycle 24. Our result is further supported by the Nobeyama radio heliograph observations which shows a smaller but statistically significant increase in prominence eruption per sunspot number around 2004 \citep{Petrie2013}. The streamers structures located between active regions and open coronal holes would tend to be weaker due to the weakening of polar fields and therefore they may be unable to prevent the eruptions of active regions and prominences. With different possible explanations, further study is required to understand the primary and secondary causes of enhanced rate of CMEs per sunspot number during a weaker sunspot cycle 24.

We have emphasized in Section~\ref{cme_or_ss} that the statistics of CMEs in the CDAW catalog depends on personal judgments and experience of the observer for identifying and counting the number CMEs. In contrast to manual CDAW catalog, the automated catalogs (e.g., CACTus, ARTEMIS, and SEEDS) use software algorithms without any human interference to identify a CME and thus the compiled CME statistics is more objective. Comparing the manual CDAW and automated CACTus catalogs during solar cycle 23, it has been suggested that CDAW catalog might have missed several narrow CMEs, especially during the maximum of the solar cycles, while the CACTus might have listed several false events as it tracks all bursty outward moving features \citep{Yashiro2008a,Robbrecht2009a}. However, there is a difference in CME statistics among the automated catalogs as they use different computer algorithms \citep{Yashiro2008a}. In this sense, there is no one-to-one correspondence between automated catalogs and they are also subjective to some extent. We admit that estimating CME occurrence rate is highly uncertain due to inclusion or exclusion of narrow CMEs in different catalogs \citep{Yashiro2008a}. However, the relative increase in CME mass loss rate seems too large to be explained by marginal detection changes. We suggest that there is a true physical increase in CME activity since the middle of solar cycle 23 and it persists during cycle 24.

The inconsistency in the rate of CMEs and ICMEs is possible as the CMEs launched at all the latitudes have been observed near the Sun while the estimates of ICMEs rate are based on the observations by a single-spacecraft located close to the ecliptic plane \citep{Richardson1995,Richardson2010,Kilpua2011}. Further, it is difficult to objectively identify the ICMEs, and therefore ICME catalogs differ in the listed number of events \citep{Richardson2010}. The inconsistency in CMEs and ICMEs rate can be possible due to the solar cycle variations in the distribution of CME source locations at the Sun as well as deflection of CMEs in the interplanetary medium. To precisely understand the divergence between the rate of CMEs and ICMEs, further study is required by investigating the location and configuration of the coronal streamer belt region, fraction of CMEs from active and polar crown filaments and the strength of polar coronal field during the rising, maximum and declining phases of the solar cycles \citep{Gopalswamy2003,Kilpua2009}.

We admit that in our study of CME mass loss rate, we have not considered the full halo CMEs as the estimation of their latitude and mass is very uncertain. Although they are around 4\% of the total number of CDAW CMEs making them statistically insignificant, they may account for a slightly larger fraction of total CME mass loss as they are highly massive and energetic \citep{Zhao2002,Gopalswamy2007}.  Further, we point out that the estimated latitudes of the CMEs are not the heliographic latitudes of the CME source regions, but they are the apparent latitudes projected on the plane of the sky. Also, the CMEs for which the CDAW CME catalog does not list their mass, we assumed their representative mass to be equal to yearly averaged CMEs mass. Such an assumption may overestimate the mass loss rate and bring uncertainties to our study when a large fraction of CMEs during a particular year are listed without their mass estimates in the catalog. It is noted that we have fitted the data points of mass loss rate and proxies of solar variability for the complete duration of the solar cycle (see, Figure~\ref{CME_massloss_ss_SC2324}). However, fitting the data points of a solar cycle by grouping them into several bins with an equal number of points could give different statistics. Thus, we admit the presence of uncertainties in the statistics due to the fitting procedure adopted in the study.

In our study, we have not investigated the phase difference between the CME rate, ICME rate, sunspot number, X-ray luminosity, and mass loss rate, however, few earlier studies have taken this into account \citep{Webb1994,Robbrecht2009a}. It would also be interesting to examine the rate of different classes of solar flares over solar cycles 23 and 24. Although our study used the non-flare X-ray background luminosity, however, it may be dominated by post-flare emissions. The relative contribution of active regions loops, quite corona and coronal holes to X-ray luminosity should also be explored over solar cycles. Our study focused on the long-term relationship among the CME mass loss rate, sunspot number, and X-ray luminosity. However, different results may be found in short or intermediate time scales. The study of the latitudinal dependence of mass loss rate and consequent torque applied on a star is important as the mass loss from higher latitudes would allow the star to maintain its high level of activity and high CME rate while losing a large amount of mass.

\subsection{Solar Wind Mass Loss Rate and Proxies of Solar Cycle}

We studied the mass loss rate via solar wind and its relationship with the solar cycle variation of sunspot number and 
X-ray background luminosity. The study shows that solar wind mass loss rate is not much dependent on the sunspot number and solar X-ray background luminosity. The solar wind mass loss rate is scattered around the average value of 2.1 $\times$ 10$^{-14}$ M$_{\odot}$  yr$^{-1}$ for cycle 23 and decreases by around 10\% for cycle 24 (Figure~\ref{sw_massloss_ss_SC2324} and \ref{sw_massloss_xr_SC2324}). Our result is in agreement to the study of \cite{Cohen2011} where they used in situ measurements of the solar wind by excluding the contribution of ICMEs from it and observations X-ay flux during solar cycle 23. However, in contrast to our study, \citet{Cranmer2017} suggested that the sphere-averaged solar wind mass loss rate is around 50\% larger at the maximum of the cycle. Using the solar wind data during solar cycles 21 and 22, they found that the correlation coefficient between the data and the fitting function is around 0.88. Our study finds that mass loss via solar wind is at least an order of magnitude higher than that from CMEs which is in agreement to \citet{Cranmer2017}. The sunspot number and X-ray background luminosity are modified by an order of magnitude through the solar cycle while the solar wind mass loss rate does not change even by a factor of 2. We think that our finding is not surprising as the proxies of solar activity represent the closed magnetic flux which modulates significantly throughout the solar cycle. However, the solar wind mass loss is largely governed by the open solar magnetic flux which is rather constant 
\citep{Owens2011}.

We also noted that the fractional contribution of CMEs to the solar wind mass flux closely tracks the solar cycle in the ecliptic. The contribution of CMEs to solar wind mass flux decreases at higher heliolatitudes. This is expected as the CME mass distribution near the Sun is also smaller at higher latitudes while the solar wind mass flux is assumed to be invariant over all the heliolatitudes. Further, the distribution of CME mass flux near 1 AU, irrespective of the latitudes, is lesser for solar cycle 24 than cycle 23. In the ecliptic region, the contribution of CMEs to the solar wind mass flux is found to be negligibly small during the solar minimum and contribution increased up to around 5\% at the maximum of solar cycles 23 and 24. Our result is in agreement of earlier studies 
\citet{Howard1985} and \citet{Lamy2017}.  Our study has excluded only the full halo CMEs for which the mass is underestimated and often some of their mass could be hidden behind the occulter of the coronagraphs. The study of \citet{Cranmer2017}, by excluding the CMEs labeled as ``poor'' events and the ones listed without masses in the CDAW catalog, have also found that the CMEs contribute only about 3\% of the background solar wind mass flux during the maximum of the cycle 23. However, in a few studies, the contribution of CMEs to solar wind mass flux in the ecliptic is found to be around 15\% during the maximum of the cycle and around 3\% at the minimum of the cycles \citep{Hildner1977,Jackson1993,Webb1994}. Most of the earlier studies, except \citet{Cranmer2017} and \citet{Lamy2017}, were based on the shorter period of data from different coronagraphs (e.g., Skylab, Solwind, and SMM) and thus involved different duty cycle corrections and inter-calibration of visibility function. Also, these studies might have suffered from instrument-dependent effects and chosen sample bias during different solar cycles. Thus, it is difficult to know the reason for larger estimates of CMEs mass flux contribution to the solar wind in most of the studies done during the 1970s to 1990s. It is possible that older coronagraphs having less sensitive than modern CCD-based instruments were more apt to observe only the massive CMEs. Therefore, consideration of those massive events as ``typical'' CMEs in the older studies might have overestimated the mean CME mass flux than that obtained from a distribution of strong and weak events. The other possibility that there has been a true decline in
CME mass flux from the 1970s-1980s to the 1990s-2000s can also contribute to the discrepancy noted in the relative contribution of CME mass flux \citep{Vourlidas2010}.

In general, our result shows that solar wind mass loss rate, although scattered around a value, is in opposite phase with the rise and maximum phase of the sunspot cycle. This is possible if coronal hole extending towards lower latitudes and open flux originating from the vicinity of active regions during solar maximum contributes to lesser mass flux along the ecliptic. The quasi-steady solar wind coming from the Sun in two fundamental states differ markedly in their speed and density \citep{Schwenn1990,Schwenn2006}. The basis of our study that measured solar wind mass flux at 1 AU along the ecliptic can represent the global mass flux from all over the solar surface is not perfectly valid in a real scenario. We admit that the intra cycle variations of solar wind mass loss rate for both the cycles are different and require an in-depth analysis of solar cycle variation in open and closed magnetic flux as well as relative mass flux from them \citep{Wang2000,Schrijver2003}. We understand that to remove additional bias in the study, the estimation of mass loss rate should be done using different CME catalog and in situ solar wind observations taken beyond the ecliptic. The estimation of CME mass flux at 1 AU is based on the several approximations regarding the structure of CMEs, their masses, and distributions. The estimation of the latitude of CMEs from their CPA as measured in coronagraphic images may not be perfectly valid for all the CMEs. Therefore, further study is required to assess the uncertainties arising due to several approximations made in our study of mass loss rate from the Sun. Future study should examine how accurately the obtained relation between different proxies of solar variability and solar mass loss rate can be extrapolated for other stars. The knowledge of star's mass loss rate can help us to constrain the evolutionary models of stellar mass, luminosity and rotation.

\section{Conclusions}
\label{conclu} 
Based on our study of estimating the mass loss rate via solar wind and CMEs during the solar cycles 23 and 24, we draw the following conclusions: 

\begin{itemize}

\item The solar wind mass loss rate, CME mass loss rate, X-ray background luminosity, and sunspot number during the maximum of solar cycle 24 decreases by 10\%, 15\%, 20\%, and 40\% respectively, than that during the maximum of cycle 23. The observed decrease in CME and solar wind mass loss rate from cycle 23 to 24 is not as large as for the sunspot number.  \\

\item We confirm that there is truly an increased rate of numerous less massive CMEs from relatively higher latitudes since 2003 to the period of solar cycle 24. Therefore, a strongly poor correlation of CME activity with sunspot number exists for solar cycle 24. \\

\item We established a relationship for mass loss rate as a linear function of monthly averaged sunspot number and solar X-ray background luminosity. This suggests that X-ray background luminosity is a better proxy for CME mass loss rate over the solar cycle than the sunspot number. However, the solar wind mass loss rate shows no obvious solar cycle dependency. \\

\item The solar wind mass loss rate is roughly an order of magnitude larger than the CME mass loss rate. The fractional contribution of CMEs to solar wind mass flux is around 5\% during solar maximum in the ecliptic and it is much smaller at higher heliolatitudes and/or during solar minimum.  \\ 

\end{itemize}

Since it is more difficult to observe the signatures of stellar CMEs and stellar spots than stellar X-ray flux, we think that the measured stellar X-ray background luminosity can probably be used as prediction tools for determining occurrence rate of CMEs from stars. The present study may guide us to understand the mass loss from other solar-type stars. However, the reliability of the extrapolation of the solar observations to other stars with much higher activity remains to be investigated in future studies.

\section*{Acknowledgments}
We acknowledge the SOHO/LASCO CME catalog which is generated and maintained at the CDAW Data Center by NASA and The Catholic University of America in cooperation with the Naval Research Laboratory. SOHO is a project of international cooperation between ESA and NASA. We also acknowledge the SIDC, CDAWEB and NGDC data center for providing the data of sunspots, solar wind and Solar X-ray flux. Y.W. is supported by the National Natural Science Foundation of China  (NSFC) grant Nos. 41574165, 41774178 and 41761134088. W.M. is supported by the NSFC grant No. 41750110481 and Chinese Academy of Sciences (CAS) President's International Fellowship Initiative (PIFI) grant No. 2015PE015.

%%%\bibliographystyle{mnras}
%%%\bibliography{wageesh_ref}

% Don't change these lines
\bsp	% typesetting comment
\label{lastpage}
\end{document}